\documentclass[a4paper,12pt]{article}
\pdfoutput=1
\usepackage{graphicx, rotating,amssymb,amsmath}

\ifx\pdfoutput\undefined
\usepackage[dvips,bookmarks]{hyperref}	
\else
\usepackage{hyperref}	
\fi
\hypersetup{colorlinks,bookmarksopen,bookmarksnumbered,citecolor=verdes,
linkcolor=blus,pdfstartview=FitH,urlcolor=rossos}
\def\myurl#1#2{\href{http://#1}{#2}}
\def\hhref#1{\href{http://arxiv.org/abs/#1}{#1}} 

\usepackage{multicol}
\usepackage{color}
\definecolor{rosso}{cmyk}{0,1,1,0.4}
\definecolor{rossos}{cmyk}{0,1,1,0.55}
\definecolor{rossoc}{cmyk}{0,1,1,0.2}
\definecolor{blu}{cmyk}{1,1,0,0.3}
\definecolor{blus}{cmyk}{1,1,0,0.6}
\definecolor{bluc}{cmyk}{1,1,0,0.1}
\definecolor{verde}{cmyk}{0.92,0,0.59,0.25}
\definecolor{verdec}{cmyk}{0.92,0,0.59,0.15}
\definecolor{verdes}{cmyk}{0.92,0,0.59,0.4}

\font\tenrsfs=rsfs10 at 12pt
\font\sevenrsfs=rsfs7
\font\fiversfs=rsfs5
\newfam\rsfsfam
\textfont\rsfsfam=\tenrsfs
\scriptfont\rsfsfam=\sevenrsfs
\scriptscriptfont\rsfsfam=\fiversfs
\def\mathscr#1{{\fam\rsfsfam\relax#1}}

\newcommand{\fig}[1]{~\ref{fig:#1}}
\newcommand{\eq}[1]{~{\rm (\ref{eq:#1})}}

\def\circa#1{\,\raise.3ex\hbox{$#1$\kern-.75em\lower1ex\hbox{$\sim$}}\,}

\newcommand{\beq}{\begin{equation}}
\newcommand{\eeq}{\end{equation}}

\def\circa#1{\,\raise.3ex\hbox{$#1$\kern-.75em\lower1ex\hbox{$\sim$}}\,}
\makeatletter

\def\art{\@ifnextchar[{\eart}{\oart}}
\def\eart[#1]#2#3#4#5#6{{\rm #2}, {#3 #4} {\rm (#6) #5} [{\hhref{#1}}]}
\def\hepart[#1]#2{{\rm #2, \hhref{#1}}}
\newcommand{\oart}[5]{{\rm #1}, {#2 #3} {\rm (#5) #4}}

\newcounter{alphaequation}[equation]
\def\thealphaequation{\theequation\hbox to
0.6em{\hfil\alph{alphaequation}\hfil}}
\def\eqnsystem#1{
\def\@eqnnum{{\rm (\thealphaequation)}}
\def\@@eqncr{\let\@tempa\relax \ifcase\@eqcnt \def\@tempa{& & &} \or
  \def\@tempa{& &}\or \def\@tempa{&}\fi\@tempa
  \if@eqnsw\@eqnnum\refstepcounter{alphaequation}\fi
\global\@eqnswtrue\global\@eqcnt=0\cr}
\refstepcounter{equation} \let\@currentlabel\theequation \def\@tempb{#1}
\ifx\@tempb\empty\else\label{#1}\fi
\refstepcounter{alphaequation}
\let\@currentlabel\thealphaequation
\global\@eqnswtrue\global\@eqcnt=0 \tabskip\@centering\let\\=\@eqncr
$$\halign to \displaywidth\bgroup \@eqnsel\hskip\@centering
$\displaystyle\tabskip\z@{##}$&\global\@eqcnt\@ne
\hskip2\arraycolsep\hfil${##}$\hfil& \global\@eqcnt\tw@\hskip2\arraycolsep
$\displaystyle\tabskip\z@{##}$\hfil
\tabskip\@centering&\llap{##}\tabskip\z@\cr}
\def\endeqnsystem{\@@eqncr\egroup$$\global\@ignoretrue} \makeatother

\begin{document}
\begin{center}
{ \hfill SACLAY--T09/030}
\color{black}
\vspace{0.3cm}

{\Huge\bf Inverse Compton constraints\\[2mm] on the Dark Matter $e^\pm$ excesses}

\medskip
\bigskip\color{black}\vspace{0.6cm}

{
{\large\bf Marco Cirelli}$^a$,
{\large\bf Paolo Panci}$^{a,b,c}$
}
\\[7mm]
{\it $^a$ Institut de Physique Th\'eorique, CNRS, URA 2306 \& CEA/Saclay,\\ 
	F-91191 Gif-sur-Yvette, France}\\[3mm]
{\it $^b$ Dipartimento di Fisica, Universit\`a degli Studi dell'Aquila, 67010 Coppito (AQ), Italy}\\[3mm]
{\it $^c$ Universit\'e Paris 7-Diderot, UFR de Physique,\\ B\^atiment Condorcet, 10, rue A.Domon et L.Duquet, 75205 Paris, France}
\end{center}

\bigskip

\centerline{\large\bf Abstract }
\begin{quote}
\color{black}\large
Recent results from experiments like PAMELA have pointed to excesses of $e^\pm$ in cosmic rays. If interpreted in terms of Dark Matter annihilations, they imply the existence of an abundant population of $e^\pm$ in the galactic halo at large. We consider the high energy gamma ray fluxes produced by Inverse Compton scattering of interstellar photons on such $e^\pm$, and compare them with the available data from EGRET and some preliminary data from FERMI. We consider different observation regions of the sky and a range of DM masses, annihilation channels and DM profiles. We find that large portions of the parameter space are excluded, in particular for DM masses larger than 1 TeV, for leptonic annihilation channels and for benchmark Einasto or NFW profiles. 

\end{quote}


\section{Introduction}
\label{intro}

While compelling evidence for the existence of Dark Matter (DM) now comes from a number of astrophysical and cosmological probes (such as galaxy rotation curves, weak lensing measurements and the precise data from the Cosmic Microwave Background observations and the Large Scale Structure surveys of the Universe~\cite{reviews}), no explicit detection has been confirmed yet. The indirect detection strategy relies on the possibility of seeing signals of the presence of DM in terms of the final products ($e^\pm$, $p$, $d$, $\gamma$, $\nu \dots$) of DM annihilations in the galactic halo, on top of the ordinary cosmic rays. 

The recent positive results from a number of indirect detection experiments have suggested the possibility that indeed such a signal has been seen. In particular, the signals point to an excess of electrons and positrons. The PAMELA satellite~\cite{PAMELA} has reported a significant excess above the expected smooth astrophysical background and a steep rise of the positron fraction $e^+/(e^++e^-)$ above 10~GeV up to 100~GeV~\cite{PAMELApositrons}, compatibly with previous less certain hints from HEAT~\cite{HEAT} and AMS-01~\cite{AMS01}.
At the same time, no signal in the $\bar p$ fluxes is seen, up to the maximal probed energy of about 100 GeV~\cite{PAMELApbar}, posing stringent constraints on Dark Matter models. 
Indeed, as discussed in detail in~\cite{CKRS}, these results, if interpreted in terms of DM annihilations, imply either
\begin{itemize}
\item[(i)] a DM particle of any mass (above about 100 GeV) that annihilates only into leptons (${\rm DM}\, {\rm DM} \to e^+e^-,\mu^+\mu^-,\tau^+\tau^-$) or
\item[(ii)] a DM particle with a mass around or above a few TeV, that can annihilate into any channel (i.e. ${\rm DM}\, {\rm DM} \to W^+W^-, ZZ, b\bar b, t\bar t$, light quark pairs and the leptonic channels above) possibly producing anti-proton fluxes at energies above those currently probed by PAMELA.
\end{itemize}
In any case, a very large annihilation cross section is needed: of the order of $10^{-23} {\rm cm}^3/{\rm sec}$ up to $10^{-20} \,{\rm cm}^3/{\rm sec}$ or more, depending on the mass of the candidate and the annihilation channel~\cite{CKRS}. These numbers are much larger than the typical cross section required by DM thermal production in cosmology ($\sim 3\cdot 10^{-26}\, {\rm cm}^3/{\rm sec}$~\cite{reviews}). They can be justified in specific models in terms of some enhancement mechanism which is effective today but not in the early universe, such as a resonance~\cite{CKRS,resonance} or Sommerfeld (see~\cite{Sommerfeld,MDMastro, CKRS}, and then~\cite{Arkani,Sommerfeld2}) enhancement, the presence of an astrophysical boost factor due to DM substructures (unlikely\,\cite{Lavalle}), or a combination of these (see e.g.~\cite{bovy}). For our purposes, they are just an input required by data.

\medskip

An excess in the flux of $e^++e^-$ has also been reported by the ATIC~\cite{ATIC-2} and PPB~\cite{PPB-BETS} balloon experiments at about 500-800 GeV. The HESS \v Cerenkov telescope, too, has published data~\cite{HESSleptons} in the range of energy from 600 GeV up to a few TeV, showing a steepening of the spectrum which is compatible both with the ATIC peak (which cannot however be fully tested) and with a power law with index $-3.05 \pm 0.02$ and a cutoff at $\approx$2 TeV. 
If interpreted in terms of DM annihilations, the ATIC peak would clearly pin down the DM mass at about 1 TeV, and as a consequence select the possibility (i) above~\cite{CKRS}. 
These balloon and \v Cerenkov signals will soon be checked by the FERMI satellite that, despite its main mission as a gamma ray telescope, will be able to measure $e^++e^-$ with unprecedented statistics~\cite{FERMIleptons}.\,\footnote{{\it Note added:} The FERMI and the HESS collaborations have published new data in~\cite{FERMIleptons2} and~\cite{HESSleptons2} after this work was completed. Their inclusion in the DM fits, replacing the ATIC data with which they are in tension, does not modifies much, however, the preferred DM green and yellow regions in fig.~\ref{fig:exclusion} (see e.g.~\cite{StrumiaPapucci}). Indeed: (i) such new data are in agreement with the PAMELA results and (ii) the HESS points show indications for a smooth cutoff that still pins down the DM mass at around a few TeV. Our analysis here remains therefore essentially unchanged.}

\bigskip

Of course, the origin of these excesses could simply lie in ordinary (albeit possibly peculiar) astrophysical sources, such as one or more pulsars~\cite{pulsars}, sources of CR in galactic spiral arms~\cite{Piran}, aged SuperNova remnants~\cite{Blasi} or exploding stars~\cite{exploding}. In this case, the sources would be located in the galactic disk, and moreover not too far from the Earth, since $e^\pm$ quickly loose energy when travelling from more that about 1 kpc away. These explanations will be confirmed or ruled out by further, more precise measurements of the spectra and possibly improved computations of the expected yields.

\bigskip

If instead Dark Matter annihilations are at the origin of the excesses, then an abundant population of high energy $e^\pm$ is being created everywhere in the DM galactic halo. In this case, an inevitable associated signal consists of high energy gamma rays (from tens of MeV up to the DM mass, which can reach several TeV) produced by the Inverse Compton Scattering (ICS, reviewed below) of the photons of the diffuse radiation fields in the galaxy (the CMB, the InfraRed -IR- background and the optical starlight -SL-) on these energetic $e^\pm$.

These gamma ray signals have to be compared with the results of the EGRET satellite, that published data relative to several windows of observation of the galactic diffuse $\gamma$-rays. Perhaps more importantly, the FERMI satellite has started publishing new results~\cite{FERMIdiffuse}, checking in particular the EGRET findings and improving very significantly the statistics.

The aim of this work is to see whether these existing and upcoming gamma ray observations have the power to test or constrain the DM interpretation on the basis of the expected associated ICS radiation.

\bigskip

In terms of their relation to DM annihilations, the ICS signals have been previously considered e.g. in~\cite{700+,Zhang,Cuoco,Barger}, in specific cases and specific DM models. Our approach differs however on a number of points: we carry out the analysis in a DM-model independent way (by considering a set of primary DM annihilation channels, scanning a wide range of DM masses and addressing the astrophysical uncertainties, e.g. varying the DM galactic distribution profiles), we consider several observational areas/datasets, and we perform most of the computations ourselves in a semi-analytical way, as opposed to running dedicated codes such as GALPROP~\cite{Galprop}. In particular, this last point implies that (a) we adopt the simplifying assumption of neglecting the diffusion processes of $e^\pm$ 
and (b) we adopt simplifying analytical approximations for the density fields of IR and optical light, modeled on the detailed results in the literature (see the discussions below). While the full numerical approach is of course more precise and probably time effective~\cite{StrumiaPapucci}, our custom semi-analytical computation can be quickly adapted to study a number of cases and allow us more control on the DM input models. When possible, we have validated our results with the results obtained with GALPROP in~\cite{700+,Cuoco}, finding a good agreement.

\bigskip

The rest of the paper is organized as follows. In Sec.~\ref{basics} we review the details of the computation of the ICS signal on the basis of the DM ingredients. In Sec.~\ref{application} we briefly describe the regions of the sky on which we focus our analysis, the corresponding observational datasets and the implementation of the formalism discussed before. In Sec.~\ref{results} we present and discuss our results and their implications for DM. Fig.\fig{exclusion} shows our main results. In Sec.~\ref{concl} we draw our conclusions.

\section{Basics of Inverse Compton Scattering computations}
\label{basics}

The Inverse Compton Scattering process consists of low energy photons that upscatter in energy on high energy electrons and positrons (as opposed to the conventional Compton Scattering of high energy photons on electrons at rest~\cite{Compton}). A pedagogical review is found in ref.~\cite{reviewICS}, on which we will base our discussion and notations. 

\medskip

In the present case, the high energy $e^\pm$ are produced by DM annihilations in any given point of the galaxy, with a density determined by the DM distribution profile and with a spectrum dictated by the primary annihilation products (as discussed below). The bath of low energy target photons consists of three main contributions: (i) the starlight originating from stars of the galactic disk (at optical wavelengths), (ii) the infrared radiation produced by the absorption and subsequent re-emission of such starlight by the galactic dust and finally (iii) the ubiquitous microwave photons of the CMB. 

\medskip

We will review in the following subsection some basic details of the computation of the ICS signal. The uninterested reader finds an executive summary in sec.~\ref{summary} and the practical details of our implementation of such formalism in sec.~\ref{application}.

\subsection{Derivation of the ICS gamma-ray spectrum}
\label{derivation}

We want to compute the differential flux $d\Phi/d\epsilon_1$ of high energy photons produced by the ICS processes, coming from an angular region of the sky denoted by $\Delta \Omega$. Here and in the following we will denote the energy of the scattered photon with $\epsilon_1$ and the energy of the original photon as $\epsilon$, in the reference system of the photon gas.  Such flux is determined as
\beq
\frac{d\Phi}{d\epsilon_1} = \frac{1}{\epsilon_1} \int_{\Delta \Omega} d\Omega \int_{\rm line-of-sight} ds\, \frac{j(\epsilon_1,r(s))}{4\pi}
\eeq
in terms of the emissivity $j(\epsilon_1,r)$ of a cell located at a distance $r$ from the galactic center. The coordinate $s$ runs along the line of sight joining the observer at Earth with the point $r$. In general, for any radiative process, the emissivity is obtained by a convolution of the spatial density of the emitting medium with the power that it radiates~\cite{Rybicki}. In our case then 
\beq
j(\epsilon_1,r)=2\int_{m_e}^{M_{\rm DM}}dE\ \mathcal{P}(\epsilon_1,E,r)\ n_e(r,E),
\eeq
where $\mathcal{P}(\epsilon_1,E, r)$ is the differential power emitted into photons of energy $\epsilon_1$ by an electron with energy $E$ and $n_e(r,E)$ is the number density at $r$ of electrons with such energy. The minimal and maximal energies of the electrons are determined by the electron mass $m_e$ and the mass of the annihilated DM particle $M_{\rm DM}$. The overall factor of 2 takes into account the fact that, beside the electrons, an equal population of positrons is produced by DM annihilations and radiates. 

The quantity $n_e(r,E)$, in full generality, has to be determined by solving the diffusion-loss equation in each point of the galaxy~\cite{reviewCR, UllioRegis} 
\beq
\underbrace{
-\frac{1}{r^2}\frac{\partial}{\partial r} \left[ r^2 D \frac{\partial f}{\partial r} \right]}_{\rm diffusion}
+ \underbrace{v \frac{\partial f}{\partial r}}_{\rm advection} 
- \underbrace{\frac{1}{3r^2}\frac{\partial}{\partial r}(r^2 v)p \frac{\partial f}{\partial p}}_{\rm convection} 
+ \underbrace{\frac{1}{p^2} \frac{\partial }{\partial p}\left[\dot p p^2 f\right]}_{\rm radiative\ losses}
= \underbrace{\frac{Q_e(E,r)}{4 \pi p^2}}_{\rm source},
\label{eq:diffusioneq}
\eeq
written here in terms of $f = n_e/(4\pi p^2)$ with $p$ the electron momentum. The advection (convection) term describes  the flow, of velocity $v$, towards (away from) the Black Hole at the Galactic Center. Since we will be interested in portions of the galactic halo that are well outside of the BH accretion region, of typical very small size $R_{\rm accr} \simeq 0.04$ pc, we assume that these advection and convection terms can be neglected. 
We neglect the diffusion term as well, in order to be able to solve eq.~(\ref{eq:diffusioneq}) in a semianalytic way (see below). This is a more drastic approximation since (for the electron energies that we consider) the characteristic time due to radiative losses can be of the order of $10^{12}$ seconds~\cite{UllioRegis}, a time in which relativistic electrons can diffuse over a distance of about one kpc comparable to the size of the smaller regions that we will consider. Diffusion could therefore redistribute somewhat the population of electrons and positrons that produce the ICS signal. We will verify {\em a posteriori} (see Sec.\ref{results}) the quantitative impact of this effect and the importance of our approximation.
Finally, in the radiative loss term we include only the ICS process. In principle all processes should be included: synchrotron radiation, bremsstrahlung, ionization and ICS. However, for relativistic electrons and for the typical magnetic field that are present in the galactic halo ($B\simeq 1\, \mu G$), ICS turns out to be the dominant one~\cite{UllioRegis}. In short: the emission due to ICS is the dominant process suffered by the electrons and positrons produced by DM annihilations, and we assume that they do not diffuse away from the production point before having radiated away most of their energy. 
In this regime, the equation is simply solved by 
\beq
n_e(r,E)=\frac1{\dot{\mathcal E}(E,r)}\int_{E}^{M_{\rm DM}}d\tilde E\ Q_e(\tilde E, r),
\eeq
where $\dot{\mathcal E}(E,r)$, related to the $\dot p$ of eq.\eq{diffusioneq}, denotes the total (i.e. into photons of any energy, not to be confused with $\mathcal{P}$) rate of electron energy loss due to ICS. 

The source term due to DM annihilations is simply given by 
\beq
Q_e(E,r)= \langle \sigma_{\rm ann} v \rangle \,\frac{\rho(r)^2}{2\,M_{\rm DM}^2} \,\frac{dN_e}{dE},
\eeq 
where $\langle \sigma_{\rm ann} v \rangle$ is the DM annihilation cross section, $\rho(r)$ is the galactic DM distribution profile and $dN_e/dE$ is the spectrum of the produced electrons, per annihilation. 

\medskip

The derivation of $\dot{\mathcal E}(E,r)$ and $\mathcal{P}(\epsilon_1,E,r)$ requires some effort, but is straightforward in terms of known Compton kinematics~\cite{reviewICS}. The total energy loss rate for an electron of energy $E$ is given by 
\beq
\dot{\mathcal E}(E,r) = \int\hspace{-0.3cm}\int_{(\epsilon,\epsilon_1)}(\epsilon_1-\epsilon) \frac{dN_{E,\epsilon}}{dt\,d\epsilon_1}
\label{eq:Edotdef}
\eeq
in terms of the rate of scatterings $dN_{E,\epsilon}/dtd\epsilon_1$ on photons of energy $\epsilon$ into photons of energy $\epsilon_1$, times the energy lost in a scattering $(\epsilon_1 - \epsilon)$ and integrated over all initial and final photon energies. The quantity $dN_{E,\epsilon}/dtd\epsilon_1$ is more conveniently computed by moving to the frame of the center of mass\footnote{{\it Compton scattering kinematics.} In the frame of the center of mass (where the electron is at rest and the photons appear to be incident on it) all quantities are denoted with a prime: $\epsilon'$ denotes the initial energy and $\epsilon^\prime_1$ the final one. One has 
\beq
\epsilon^\prime = \gamma \epsilon (1-\cos\theta), \qquad \epsilon^\prime_{\rm min} = \frac{\epsilon}{2\gamma}, \qquad \epsilon^\prime_{\rm max} = 2\gamma \epsilon,
\eeq
where $\theta$ is the approach angle of a photon in the photon gas system and $\gamma \gg 1$ is the Lorentz factor. One also has the well-known relation
\beq
\epsilon_1^\prime=\frac{\epsilon^\prime}{1+\frac{\epsilon^\prime}m\left(1-\cos\theta_1^\prime\right)},
\eeq
where $\theta_1^\prime$ is the scattering angle in the center of mass. The relation between the final energies in the two frames reads
\beq
\epsilon_1 = \gamma \epsilon_1^\prime (1-\cos\theta_1^\prime).
\eeq
The limit of Compton scattering in which the energy of the photon before scattering in the electron rest frame is much less than the electron mass defines the {\em Thomson limit}. In this case the final energy of the photon, in the same frame, is unchanged (the electron looses a negligible fraction of its energy at each scattering). Back in the photon gas frame, these conditions translate in the fact that the initial photon energy is upscattered to a maximum energy that is larger by a factor $4 \gamma^2$. In formul\ae 
\beq
\epsilon^\prime \ll m, \qquad \epsilon_1^\prime \simeq \epsilon^\prime, \qquad  \epsilon_1^{\rm max} = 4 \epsilon \gamma^2 \qquad [{\rm Thomson\ limit}].
\label{Thomsonlimit}
\eeq
It is easy to check that, for typical electrons of DM origin ($E \lesssim 1$ TeV, $\gamma \lesssim 2\,10^6$) the Thomson limit is verified for the scattering of CMB photons, since $\epsilon^\prime_{\rm max} = 2\, \gamma\, \epsilon_{\rm CMB} \lesssim 100\ {\rm eV} \ll m$, but {\it not} for starlight photons of typical temperature $0.3$ eV, for which $\epsilon^\prime_{\rm max} \simeq 1$ MeV.
} 
as
\beq
\frac{dN_{E,\epsilon}}{dt\,d\epsilon_1}=\int \hspace{-0.3cm} \int_{(\Omega_1^\prime,\epsilon^\prime)}\frac{dN_{E,\epsilon}}{dt^\prime\,d\epsilon_1^\prime\,d\Omega_1^\prime\,d\epsilon^\prime}\frac{dt^\prime}{dt}\frac{d\epsilon_1^\prime}{d\epsilon_1}
\label{scatterings}
\eeq
with $dt^\prime/dt=1/\gamma$ and $d\epsilon_1^\prime/d\epsilon_1 \simeq1/[\gamma(1-\cos\theta_1^\prime)]$, having denoted the Lorentz factor of the electron (always assumed to be highly relativistic) as $\gamma = E/m \gg 1$.
In the new frame, the rate of scatterings is given by 
\beq
\frac{dN_{E,\epsilon}}{dt^\prime\,d\epsilon_1^\prime\,d\Omega_1^\prime\,d\epsilon^\prime} =
\frac{dn^\prime(\epsilon^\prime;\epsilon)}{d\epsilon^\prime}\frac{d\sigma}{d\epsilon_1^\prime\,d\Omega_1^\prime}.
\eeq
Here the last term is just the differential Compton cross section, and $dn^\prime(\epsilon^\prime;\epsilon)/d\epsilon^\prime$ expresses the (differential) density of photons within $d\epsilon^\prime$ due to photons that are within $d\epsilon$ in the photon gas frame. This quantity, in turn, is to be determined on the basis of the density of photons in the original reference frame, which can be done by using proper relativistic invariants. One obtains~\cite{reviewICS} 
\beq
dn^\prime(\epsilon^\prime;\epsilon)\, d\epsilon^\prime = \frac{1}{2}n(\epsilon,r)\ d\epsilon\ d(\cos\theta)\, \frac{\epsilon^\prime}{\epsilon},
\eeq
where finally $n(\epsilon,r)$ is indeed the photon distribution in the photon gas frame. The last formula holds if one neglects possible anisotropies in the spatial distribution of the target photons. This is of course precisely true only for CMB photons while it is an approximation for starlight and infrared photons. 
 
The Compton cross section is given, in full generality, by the Klein-Nishina formula
\beq
\frac{d\sigma}{d\epsilon_1^\prime d\Omega_1^\prime } = \frac{3}{16 \pi} \sigma_{\rm T} \left( \frac{\epsilon_1^\prime}{\epsilon^\prime} \right)^2 \left( \frac{\epsilon^\prime}{\epsilon_1^\prime} + \frac{\epsilon_1^\prime}{\epsilon^\prime} - \sin^2\theta^\prime_1 \right) \delta \Big(\epsilon_1^\prime - \frac{\epsilon^\prime}{1+\frac{\epsilon^\prime}{m}(1-\cos\theta_1^\prime)}\Big),
\eeq
where $\sigma_{\rm T} = 8 \pi r_e^2/3 = 0.6652$ barn is the total Thomson cross section in terms of the classical electron radius $r_e$. In the Thomson limit (of eq.(\ref{Thomsonlimit})), it simplifies to 
\beq
\frac{d\sigma}{d\epsilon_1^\prime\,d\Omega_1^\prime}\simeq \frac{3}{16\pi}\sigma_{\rm T}(1+\cos^2\theta_1^\prime)\delta \Big(\epsilon_1^\prime-\epsilon^\prime \Big) \qquad {\rm [Thomson\ limit]}.
\eeq

It is now possible, by solving the integrals in the electron rest frame, to compute the rate of scattering of eq.(\ref{scatterings}) as~\cite{reviewICS}
\beq
\frac{dN_{E,\epsilon}}{dt\,d\epsilon_1} = 3 \sigma_{\rm T} \frac{n(\epsilon,r)d\epsilon}{4\gamma^2\epsilon}\left[ 2q\ln q+q+1-2q^2+\frac{1}{2}\frac{(\Gamma_\epsilon q)^2}{1+\Gamma_\epsilon q}(1-q) \right], 
\eeq
in terms of
\beq
q=\frac{\tilde\epsilon_1}{\Gamma_\epsilon(1-\tilde\epsilon_1)}, \qquad {\rm with}\  \Gamma_\epsilon=\frac{4\epsilon\gamma}{m},\quad \tilde\epsilon_1=\frac{\epsilon_1}{\gamma m},  \quad {\rm in} \ \frac{1}{4\gamma^2}\simeq 0 \le q \le 1. 
\eeq
In the Thomson limit, the same expression holds with $\Gamma_\epsilon \to 0$ (so that the last addend vanishes) and, since $\tilde\epsilon_1 \ll 1$, $q \to y = \frac{\epsilon_1}{4\gamma^2\epsilon}$. 
So finally, plugging this result into\eq{Edotdef} one obtains explicitly\footnote{The integral in $q$ can be done analytically, it just leads to a long expression.}
\beq
\begin{split}
& \dot{\mathcal E}(E,r) = \\
& 3\sigma_{\rm T} \int_0^\infty d\epsilon\ \epsilon \int_{1/4\gamma^2}^1 dq\ n(\epsilon,r) \frac{(4\gamma^2-\Gamma_\epsilon)q-1}{(1+\Gamma_\epsilon q)^3}\left[ 2q\ln q+q+1-2q^2+\frac{1}{2}\frac{(\Gamma_\epsilon q)^2}{1+\Gamma_\epsilon q}(1-q) \right]. 
\end{split}
\label{eq:Edot}
\eeq
In the Thomson limit it reduces to the particularly compact expression
\beq
\dot{\mathcal E}(E,r) = \frac{4}{3}\sigma_{\rm T} \gamma^2 \int_0^\infty d\epsilon\ \epsilon\ n(\epsilon,r) \qquad {\rm [Thomson\ limit]}.
\label{eq:EdotThomson}
\eeq
that shows that the total energy lost is proportional to the second power of the electron energy $E=\gamma m$ and to the energy density of the photon bath.

\medskip

The computation of $\mathcal{P}(\epsilon_1,E,r)$ proceeds in the same way, apart from the fact that no integral over $\epsilon_1$ is involved (by definition of $\mathcal{P}$). One has in fact 
\beq
\mathcal{P}(\epsilon_1,E,r) =
\int_{(\epsilon)}(\epsilon_1-\epsilon) \frac{dN_{E,\epsilon}}{dt\,d\epsilon_1}
\eeq
from which
\begin{equation}
\begin{split}
& \mathcal{P}(\epsilon_1,E,r)  = \\
&  \frac{3 \sigma_{\rm T}}{4\gamma^2} \epsilon_1 \hspace{-0.2cm} \int_{1/4\gamma^2}^1\hspace{-0.65cm} dq  \left(1-\frac{1}{4q\gamma^2 (1-\tilde\epsilon_1)} \right) \frac{n\big(\epsilon(q),r\big)}{q} \left[ 2q\ln q+q+1-2q^2+\frac{1}{2}\frac{\tilde\epsilon_1^2}{1-\tilde\epsilon_1}(1-q) \right].
\end{split}
\label{eq:power}
\end{equation}
In the Thomson limit
\beq
\mathcal{P}(\epsilon_1,E,r) = \frac{3 \sigma_{\rm T}}{4\gamma^2} \epsilon_1  \int_{0}^1 \hspace{-0.2cm} dy   \frac{n\big(\epsilon(y),r\big)}{y} \left[ 2y\ln y+y+1-2y^2 \right]\qquad {\rm [Thomson\ limit]}.
\label{eq:powerThomson}
\eeq

\subsection{Summary}
\label{summary}
With the ingredients above, the differential flux of ICS photons (of energy $\epsilon_1$) from a region $\Delta \Omega$ of the sky is given by
\beq
\frac{d\Phi}{d\epsilon_1} = \frac{1}{\epsilon_1}\frac{\langle \sigma_{\rm ann} v \rangle}{4\pi} r_\odot \frac{\rho_\odot^2}{M_{\rm DM}^2} \int_{\Delta \Omega}\hspace{-0.3cm}d\Omega \int_{\rm l.o.s.} \frac{ds}{r_\odot} \left( \frac{\rho(r)}{\rho_\odot} \right)^2 \int_m^{M_{\rm DM}}\hspace{-0.3cm}dE\,  \frac{{\mathcal P}(\epsilon_1,E,r)}{\dot{\mathcal E}(E,r)} Y(E), 
\label{eq:summary}
\eeq
where $Y(E) = \int_E^{M_{\rm DM}} d\tilde E \, \frac{dN}{d\tilde E}$ is the number of electrons generated with energy larger than $E$ in one annihilation. Here $r_\odot \simeq 8.5$ kpc is the distance of the Sun from the galactic center and $\rho_\odot \simeq 0.3\ {\rm GeV}/{\rm cm}^3$ the local normalization of the DM density.  
This expression is subject only to the assumption of highly relativistic electrons and isotropy of the target photon gas.

The total energy loss $\dot{\mathcal E}(E,r)$ is given in eq.\eq{Edot}, the power ${\mathcal P}(\epsilon_1,E,r)$ in eq.\eq{power}. In the Thomson limit (valid e.g. for TeV electrons scattering on the CMB, but not on starlight), eq.\eq{EdotThomson} and \eq{powerThomson} can be used.

\medskip

The $r$ dependence in ${\mathcal P}$ and $\dot{\mathcal E}$ comes from that of the photon bath $n(\epsilon,r)$. If it is not present (such as for the case of the uniformly distributed CMB photons) or it is neglected (as we will assume below for starlight and infrared photons), the computation of the flux simplifies considerably since the integral over $r$ can be performed separately and the result cast in terms of the usual average geometrical factor $\bar J$~\cite{Buckley}. Explicitly 
\beq
\frac{d\Phi}{d\epsilon_1} = \frac{1}{\epsilon_1}\frac{\langle \sigma_{\rm ann} v \rangle}{4\pi} r_\odot \frac{\rho_\odot^2}{M_{\rm DM}^2}  \bar J \, \Delta\Omega \int_m^{M_{\rm DM}}\hspace{-0.5cm}dE\, \frac{{\mathcal P}(\epsilon_1,E)}{\dot{\mathcal E}(E)} Y(E), \quad \bar J \, \Delta\Omega = \int_{\Delta \Omega}\hspace{-0.3cm}d\Omega \int_{\rm l.o.s.} \frac{ds}{r_\odot} \left( \frac{\rho(r)}{\rho_\odot} \right)^2.
\label{eq:summarynor}
\eeq 

\section{Datasets and Application}
\label{application}

A number of observations of the diffuse gamma ray emission are available in the literature. 
In particular, the Energetic Gamma-Ray Experiment Telescope (EGRET) instrument, operational in the 90's aboard the 
Compton Gamma Ray Observatory, has produced data on several windows sitting across the galactic plane~\cite{Hunter1997} or outside of it~\cite{Sreekumar1997}. The FERMI satellite~\cite{FERMIingen} has started exploring the same areas, with much higher sensitivity, and other regions on the sky.

\medskip

On the basis of the discussion in~\cite{Strong2004}, we consider for definiteness the following three regions (see the summary in Table~\ref{tab:regions}):
\begin{itemize}
\item A rectangular region encompassing the inner galaxy area, at galactic latitude $0.25^\circ < |b| < 4.75^\circ$ and galactic longitude $330.25^\circ < l < 359.75^\circ$ and $0.25^\circ < l < 29.75^\circ$ (denoted in the following as `5$\times$30' region); we take the EGRET data points from the reanalysis in~\cite{5x30}. They reach about 80 GeV.
\item A rectangular region larger then the previous one, at galactic latitude $0.25^\circ < |b| < 9.75^\circ$ and galactic longitude $300.25^\circ < l < 359.75^\circ$ and $0.25^\circ < l < 59.75^\circ$ (denoted in the following as `10$\times$60' region). This is the original area considered in~\cite{Hunter1997}. We take the EGRET data points from~\cite{Strong2004}. They also reach about 80 GeV.
\item The intermediate latitude region defined by $10^\circ < |b| < 20^\circ$ in galactic latitude, at all galactic longitudes (denoted in the following as `10$-$20 strips'); we take the EGRET data from~\cite{Strong2004}\footnote{The original reference~\cite{Sreekumar1997} focussed on a larger portion of the sky.}. 
The datasets in this region extend up to only $\approx$ 10 GeV in energy.

For this region we also consider the preliminary data of the FERMI satellite, on the basis of~\cite{FERMIdiffuse}. These data do not confirm the presence of the `GeV excess' reported by EGRET, but instead follow the expected purely astrophysical background. Given the preliminary nature of the FERMI results and the fact that they are currently provided for this region only, we will here show exclusion curves for both datasets. But we comment later on the implications of the possibility that FERMI does not confirm the `EGRET excess' anywhere.

\end{itemize}

\begin{table}[t]
\centering
\footnotesize{
\begin{tabular}{l|c|ccc|ccc}
Region & \multicolumn{1}{c|}{latitude $b$ \&}  &  \multicolumn{3}{c}{$\bar J $} & \multicolumn{3}{|c}{ISRF}\\
 & \multicolumn{1}{c|}{longitude $l$} &  IsoT & NFW  & Einasto & $\mathcal{N}_{\rm SL}$ & $\mathcal{N}_{\rm IR}$ & $\mathcal{N}_{\rm CMB}$ \\
\hline
\vspace{-3mm} & & & & & & & \\
`5$\times$30'   &$0.25^\circ < $ $|b|$ $< 4.75^\circ$  & 10.0 & 52.1 & 95.5 & $1.7 \cdot 10^{-11}$ & $7.0 \cdot 10^{-5}$ & 1\\
 & $330.25^\circ < $ $ l$   $< 359.75^\circ$ &  & & & & & \\
  & $0.25^\circ < $ $ l $   $< 29.75^\circ$ & & & & & & \\[1mm]
\hline 
\vspace{-3mm} & & & & & & & \\
`10$\times$60' & $0.25^\circ <$ $|b|$ $< 9.75^\circ$ & 6.5 & 21.2 & 35.8 & $2.7 \cdot 10^{-12}$ & $7.0 \cdot 10^{-5}$ & 1 \\
 & $300.25^\circ < $ $l$ $< 359.75^\circ$ & & & & & & \\
 & $0.25^\circ < $ $l$ $< 59.75^\circ$ & & & & & & \\[1mm]
\hline 
\vspace{-3mm} & & & & & & & \\
`10$-$20' & $10^\circ <$ $|b|$ $< 20^\circ$  &  2.3 & 3.4 & 4.3 & $8.9 \cdot 10^{-13}$ & $1.3 \cdot 10^{-5}$ & 1 \\
& $0^\circ < $ $l$ $ < 360^\circ$ & & & & & & \\
\hline\\[-3mm]
\multicolumn{5}{r}{$T_i=$} & 0.3 eV& 3.5 meV & 2.725 K
\end{tabular}
\caption{\em Summary of the observational regions that we consider, together with the corresponding values of the average $\bar J$ factor for different DM halo profiles and the parameters of the modelization of the Inter-Stellar Radiation Field.\label{tab:regions}}}
\end{table}

\medskip

In order to compare with these datasets the ICS signals discussed in the previous section we need to specify prescriptions for the electron spectrum $dN_e/dE$, the distribution of target photons $n(\epsilon,r)$ and the distribution of Dark Matter in the galaxy $\rho(r)$.

\medskip

We consider the electron spectra from DM annihilations for the following primary channels:
\beq
{\rm DM}\, {\rm DM}\to e^+e^-, \mu^+\mu^-, \tau^+\tau^-, W^+W^-, b\bar b\ \ {\rm and}\ \ t\bar t,
\label{eq:channels}
\eeq
computed with the use of the PYTHIA MonteCarlo code~\cite{PYTHIA} as described in detail in~\cite{CKRS}.\footnote{We stick to the case of ``direct" annihilation of DM particles into a pair of Standard Model particles, i.e. we do not study models in which the annihilation proceeds into some new light mediator state (see for instance~\cite{StrumiaPapucci, Arkani, Papucci, Nomura,BertoneTaosoSweden,Schwetz}). These ``cascade annihilation processes'' lead generically to softer electron spectra.} We scan over a large range of DM masses, from 100 GeV up to several tens of TeV.

\medskip

\begin{figure}[t]
\begin{center}
\includegraphics[width=0.485\textwidth]{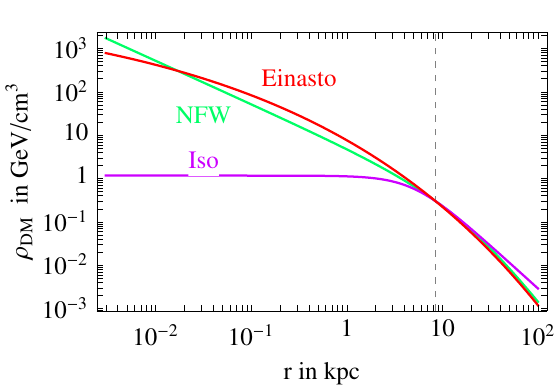}\ 
\includegraphics[width=0.50\textwidth]{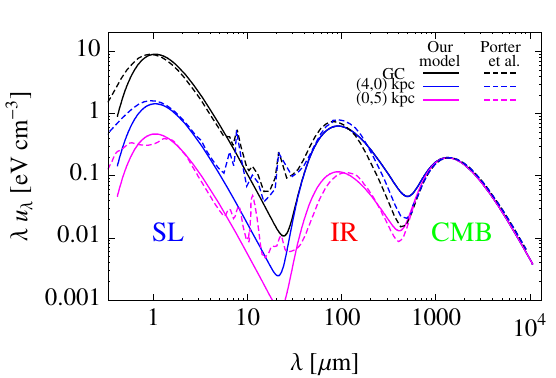}
\caption{\em\label{fig:ISRF} {\bf Left:} The Dark Matter galactic distribution profiles considered in the text. {\bf Right:} Modelization of the total Inter-Stellar Radiation Field (ISRF): the dashed lines reproduce the computations of~\cite{ISRF,ISRF2}. From top to bottom: Galactic Center ($R=0$ kpc, $z=0$ kpc -- black), $R=4$ kpc, $z=0$ kpc (blue), 5 kpc outside of the galactic plane (magenta). The solid lines represent our modelizations in terms of (renormalized) blackbody spectra for the StarLight and InfraRed photons.}
\end{center}
\end{figure}

As for the photon density distribution $n(\epsilon)$, we adopt the following prescriptions, based on the precise modelizations of the Inter-Stellar Radiation Field discussed in~\cite{ISRF, ISRF2}. First, for each region that we study we consider an {\em average} density field that does not depend on $r$ (but is different for each region, as discussed below). This allows us to use the formalism of eq.\eq{summarynor} to compute the ICS signal. Second, we choose to always model the total radiation density as a superposition of three blackbody-like spectra ($i=1,2,3$)
\beq
n_{a,i}(\epsilon) = {\mathcal N}_a\ \frac{\epsilon^2}{\pi^2}\frac{1}{(e^{\epsilon/T_i}-1)}
\eeq
with different temperatures: one for the CMB with $T_{\rm CMB}=2.753$ K (which is of course not an approximation), one for the InfraRed light with $T_{\rm IR}= 3.5 \cdot 10^{-3}$ eV and one for the StarLight with $T_{\rm SL}=0.3$ eV. Fig.\fig{ISRF} shows that this approximation is actually quite good, when compared with the precise computations, in grasping the general shape of the spectrum (except perhaps for the details at the junction wavelengths between starlight and infrared light, where the density is however suppressed). 
In order to take into account the spatial variation of the SL and IR fields of radiation throughout the galaxy, we normalize differently the blackbody spectra in each of the three regions that we study, with the overall coefficients ${\mathcal N}_a$ ($a=1,2,3$).
More precisely, for the `5$\times$30' region we take the ISRF characteristic of the Galactic Center ($R=0,z=0$, in galactic cylindrical coordinates)~\cite{ISRF}, which features an important contribution at small wavelengths of starlight photons, because of the higher density of stars in the galactic bulge.
For the `10$\times$60' region  we adopt the ISRF for $R=4\, {\rm kpc},z=0$ from~\cite{ISRF},
while for the `10$-$20' strips we use the ISRF computed for distances above the galactic plane, $z=5$\, kpc, in~\cite{ISRF2}. In the latter, the SL and IR intensity is significantly lower, as expected, but still sizable~\cite{ISRF2}. 
Table~\ref{tab:regions} summarizes the relevant parameters and fig.\fig{ISRF} illustrates the result.

\medskip

Finally, for the Dark Matter galactic profiles $\rho(r)$ we consider three different models determined by numerical simulations. Recent, state-of-the-art computations seem to converge towards the so called Einasto profile~\cite{Graham:2005xx, Navarro:2008kc}
\begin{equation}
 \rho_{\rm Ein}(r)=\rho_{s}\exp\left[-\frac{2}{\alpha}\left(\left(\frac{r}{r_{s}}\right)^{\alpha}-1\right)\right],\ \ \alpha=0.17.
   \label{eq:Einasto}
\end{equation}
The Navarro-Frenck-White profile~\cite{Navarro:1995iw} and the isothermal profile~\cite{Bahcall:1980fb}\\[3mm]
\begin{minipage}{0.495\linewidth}
\begin{equation}
 \rho_{\rm NFW}(r)=\rho_{s}\frac{r_{s}}{r}\left(1+\frac{r}{r_{s}}\right)^{-2}
   \label{eq:NFW}
\end{equation}
\end{minipage}
\vspace{3mm}
\begin{minipage}{0.495\linewidth}
\vspace{2mm}
\begin{equation} 
   \rho_{\rm isoT}(r)=\frac{\rho_{s}}{1+\left(r/r_{s}\right)^{2}}
   \label{eq:isoT}
\end{equation}
\end{minipage}
represent instead previously standard choices. The values for the parameters $r_{s}$ and $\rho_{s}$ of the three models are given by 
$$ \begin{tabular}{l|cc}
  DM halo model & $r_{s}$ in kpc & $\rho_{s}$ in GeV/cm$^{3}$\\
  \hline
  NFW \cite{Navarro:1995iw} & 20 & 0.26\\
  Einasto \cite{Graham:2005xx, Navarro:2008kc} & 20 & 0.06\\
  Isothermal \cite{Bahcall:1980fb} & 5 & 1.16
 \end{tabular}$$
(note that all profiles are normalized at $\rho_\odot$ at the location of the Earth). They are plotted in fig.~\ref{fig:ISRF}. 
These profiles sensibly differ at the Galactic Center (from the cored isothermal profile to the more cuspy NFW). They are however also appreciably different a few kpc away from it. Since we consider integrated signals from regions of observation that extend along several kpc from the GC, a dependence on the choice of profile is therefore to be expected.

\section{Results}
\label{results}

\begin{figure}[t]
\begin{center}
\hspace{-0.5cm}
\includegraphics[width=0.335\textwidth]{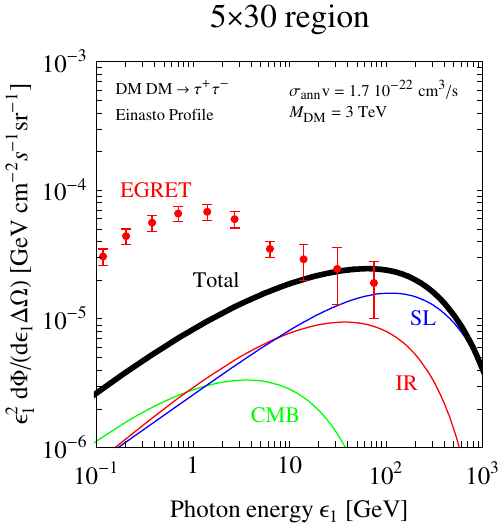}\
\includegraphics[width=0.335\textwidth]{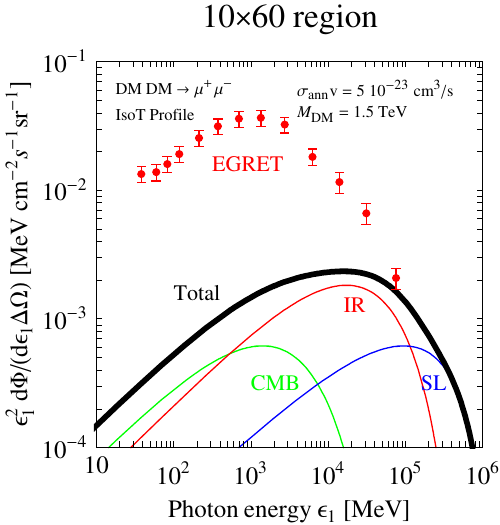}\
\includegraphics[width=0.335\textwidth]{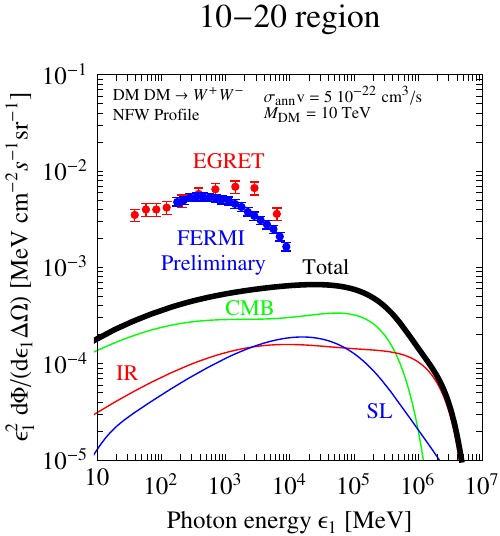}
\caption{\em\label{fig:signals} Some examples of ICS signals from selected DM models, in the three different regions of the sky that we consider, superimposed to the relevant datasets. 
{\bf Left}: in the region `5$\times$30', EGRET datapoints and the signal from a $3$ TeV DM candidate annihilating into $\tau^+\tau^-$ with $\sigma_{\rm ann}v = 1.7 \cdot 10^{-22}\, {\rm cm}^3/{\rm sec}$, choosing the Einasto DM profile of eq.\eq{Einasto} (in analogy with~\cite{700+}, where however the Einasto profile differs slightly). 
{\bf Center}: in the region `10$\times$60', EGRET datapoints and a signal from a $1.5$ TeV DM candidate annihilating into $\mu^+\mu^-$ with $\sigma_{\rm ann}v = 5 \cdot 10^{-23}\, {\rm cm}^3/{\rm sec}$, choosing the isothermal DM profile of eq.\eq{isoT} (in analogy with~\cite{Cuoco}, where however the isothermal profile differs slightly). 
{\bf Right}: in the region `10$-$20' strips, EGRET datapoints and preliminary FERMI datapoints, together with a signal from a $10$ TeV DM candidate annihilating into $W^+W^-$ with $\sigma_{\rm ann}v = 5 \cdot 10^{-22}\, {\rm cm}^3/{\rm sec}$, choosing the NFW DM profile of eq.\eq{NFW}.}
\end{center}
\end{figure}

\begin{figure}[p]
\begin{center}
\hspace{-8mm}
\includegraphics[width=0.333\textwidth]{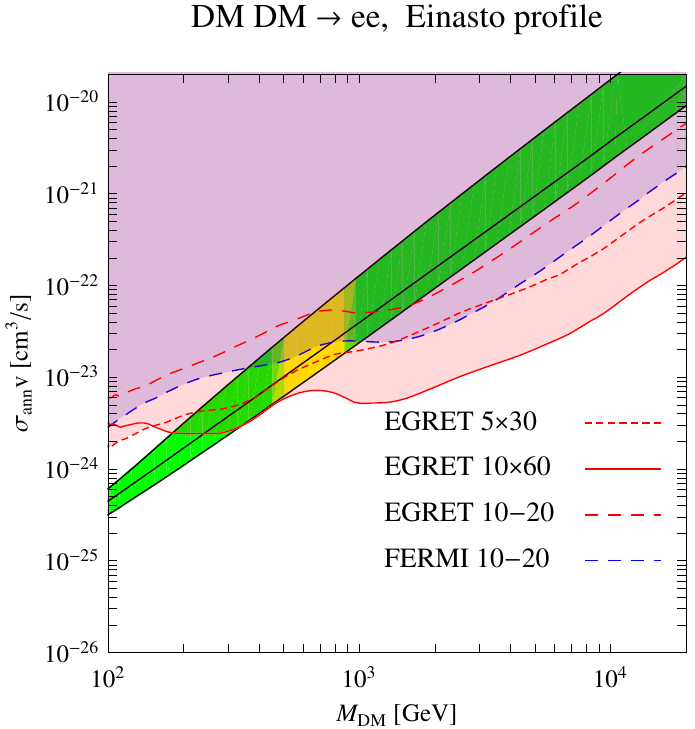}\
\includegraphics[width=0.333\textwidth]{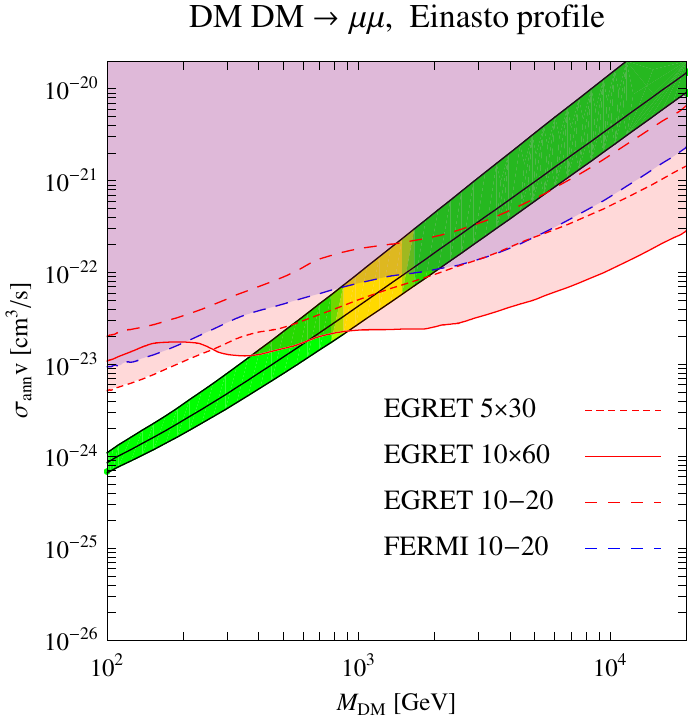}\
\includegraphics[width=0.333\textwidth]{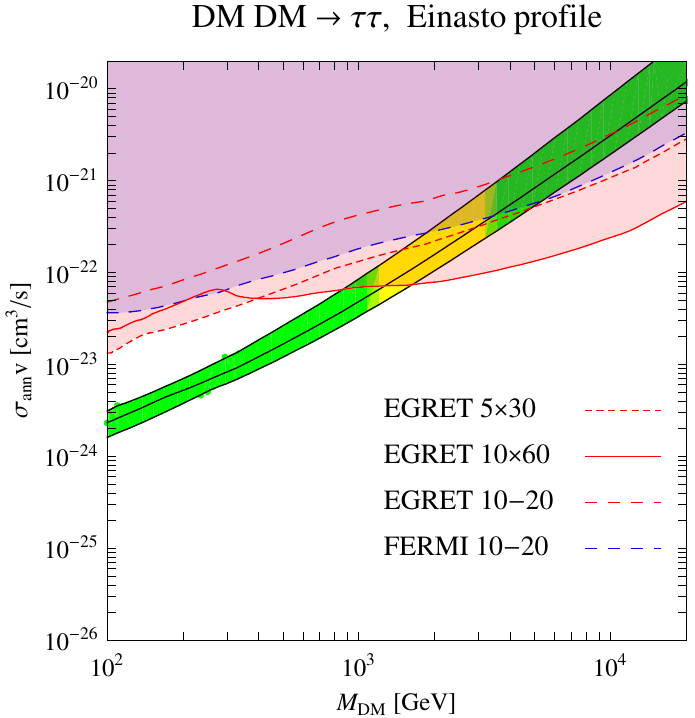}\\[2mm]
\hspace{-8mm}
\includegraphics[width=0.333\textwidth]{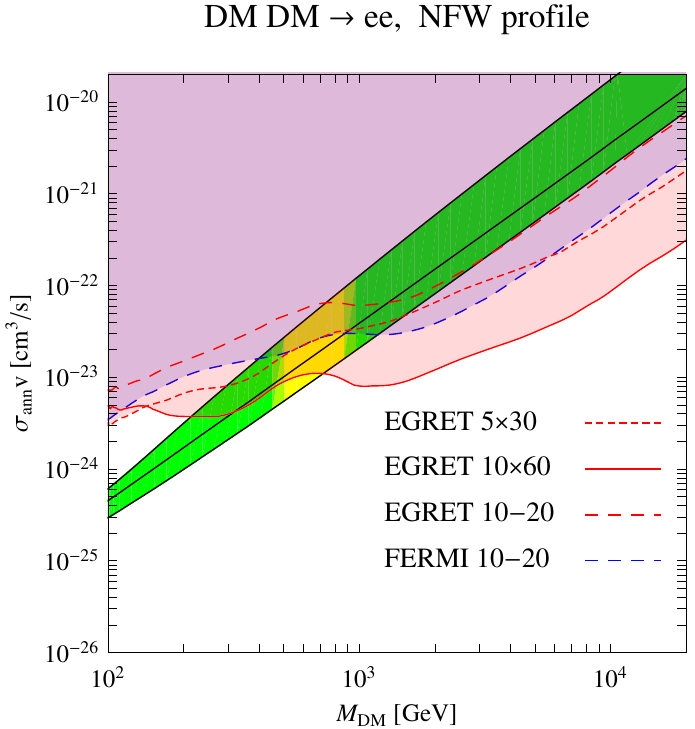}\
\includegraphics[width=0.333\textwidth]{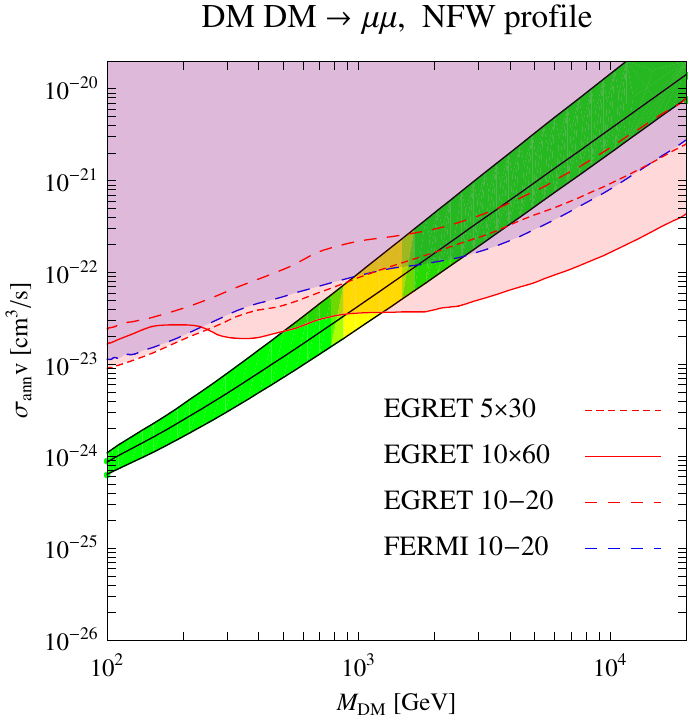}\
\includegraphics[width=0.333\textwidth]{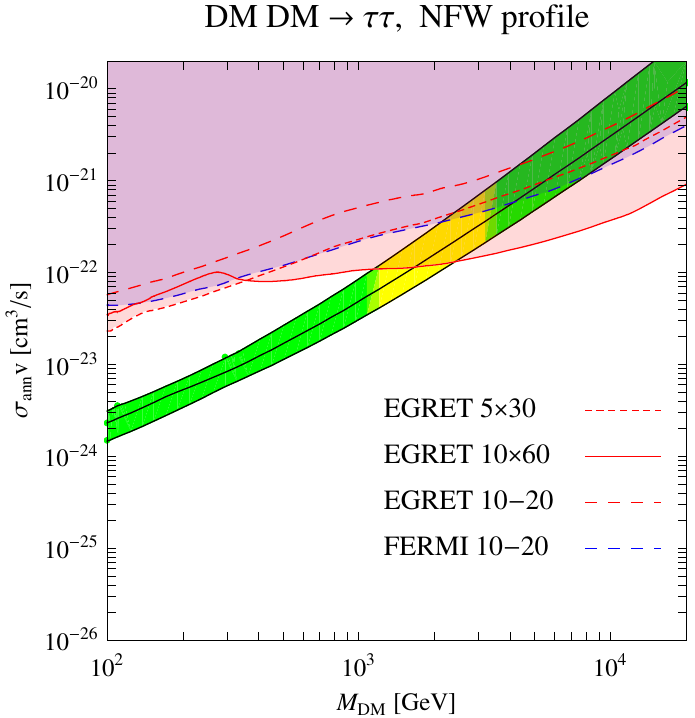}\\[2mm]
\hspace{-8mm}
\includegraphics[width=0.333\textwidth]{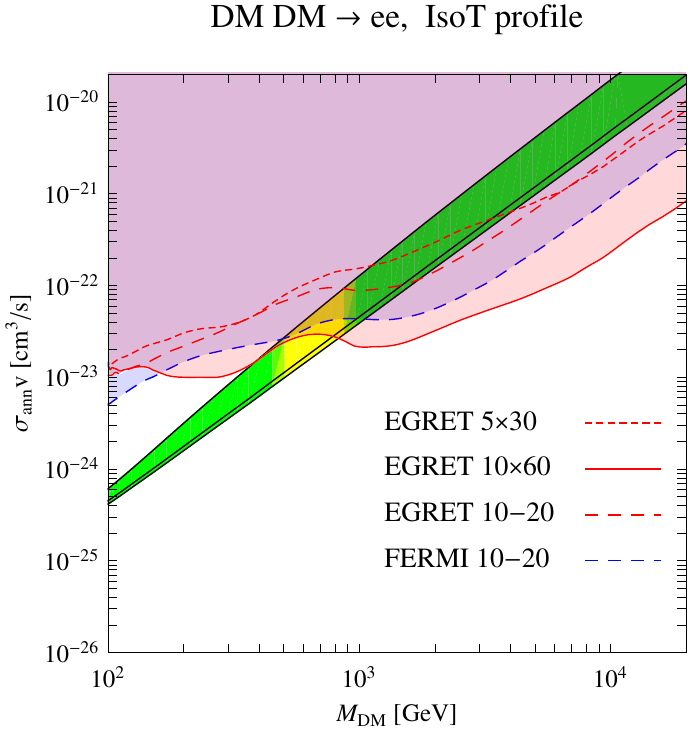}\
\includegraphics[width=0.333\textwidth]{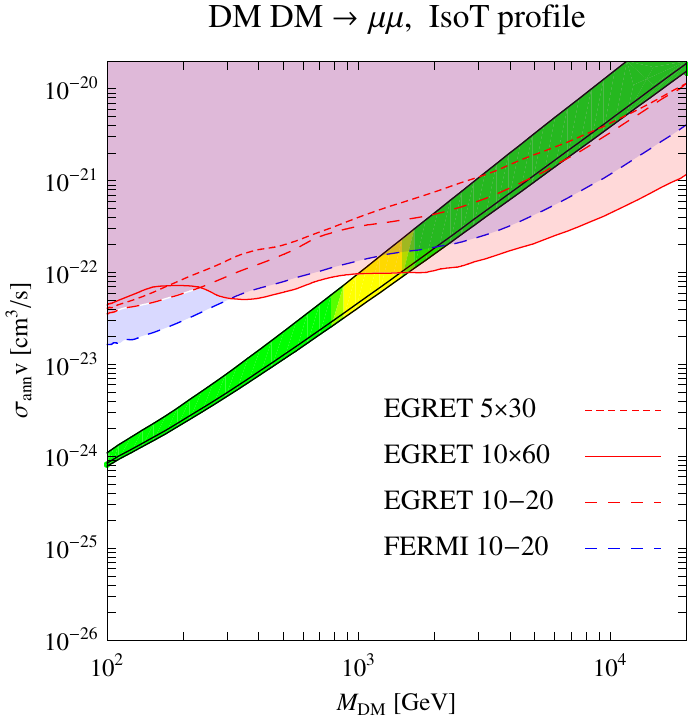}\
\includegraphics[width=0.333\textwidth]{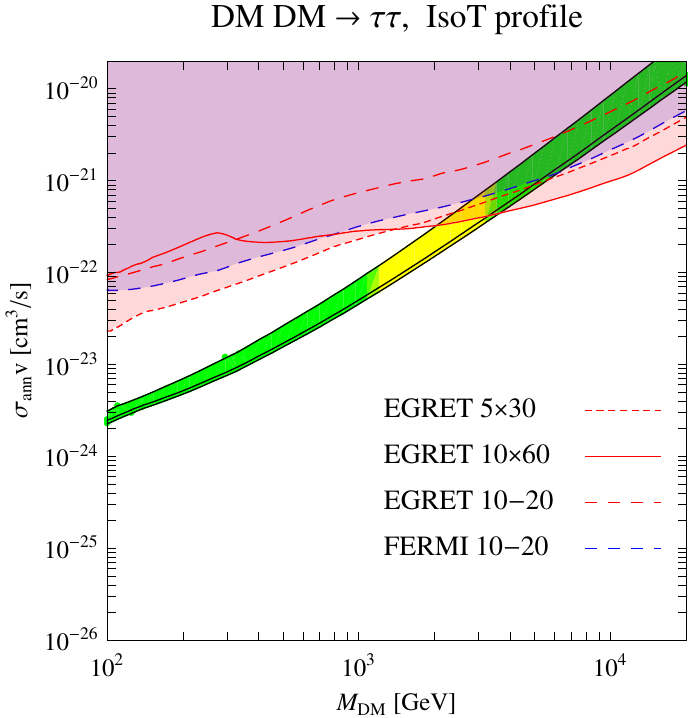}
\caption{\em\label{fig:exclusion} The regions favored by PAMELA (green bands), containing in particular the areas favored by PAMELA+ATIC (yellow areas), are compared with the bounds from ICS secondary radiation. The first column of panels refers to DM annihilations into $e^+e^-$, the second into $\mu^+\mu^-$ and the third into $\tau^+\tau^-$; the three rows assume respectively an Einasto, an NFW and an isothermal profile. In each panel, the bounds from EGRET data in the `5$\times$30' region are plotted with a short dashed red line, those from EGRET data in the `10$\times$60' region with a solid red line and those from EGRET data in the `10$-$20' strips with a dashed red line. The preliminary FERMI bounds in the `10$-$20' strips are plotted with a dashed blue line. The ICS bounds are computed adopting the simplifying assumption of neglecting the diffusion of the source $e^\pm$, see text for the full discussion.}
\end{center}
\end{figure}

\begin{figure}[p]
\begin{center}
\hspace{-8mm}
\includegraphics[width=0.333\textwidth]{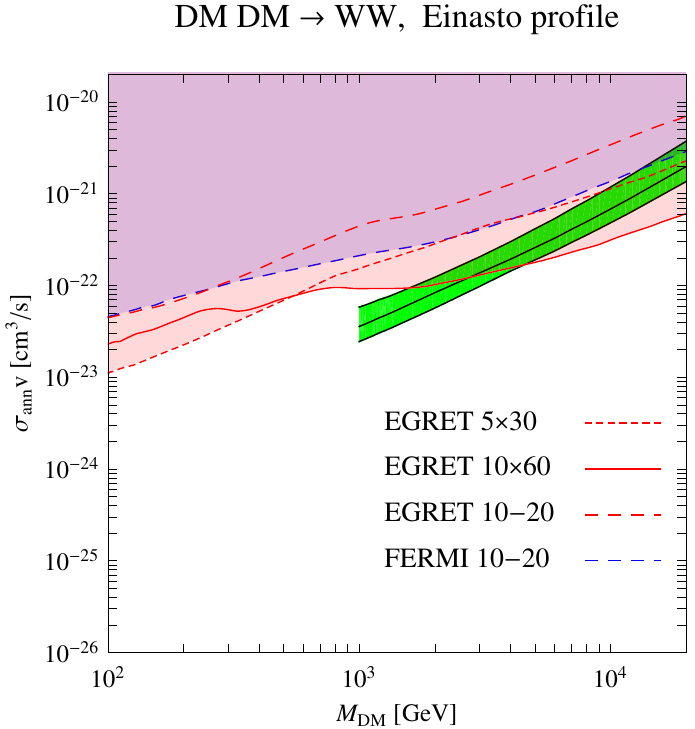}\
\includegraphics[width=0.333\textwidth]{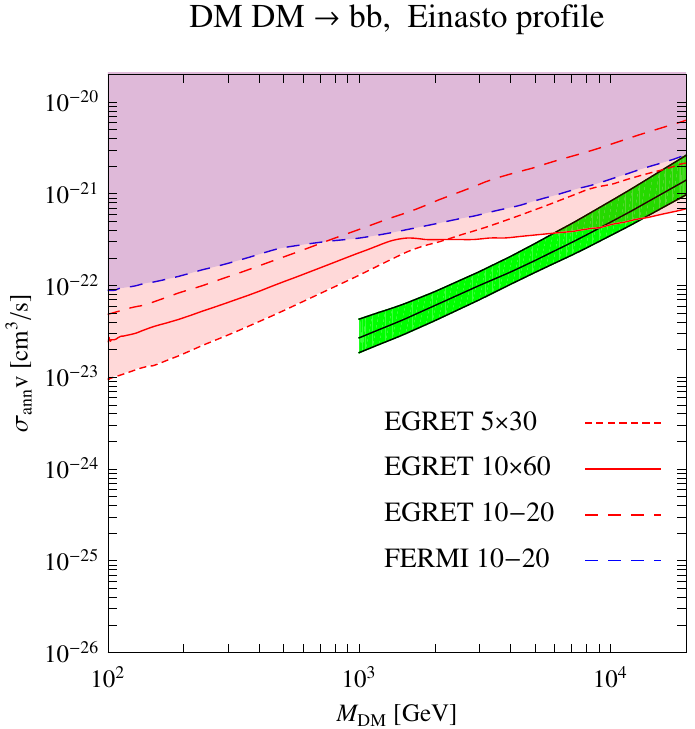}\
\includegraphics[width=0.333\textwidth]{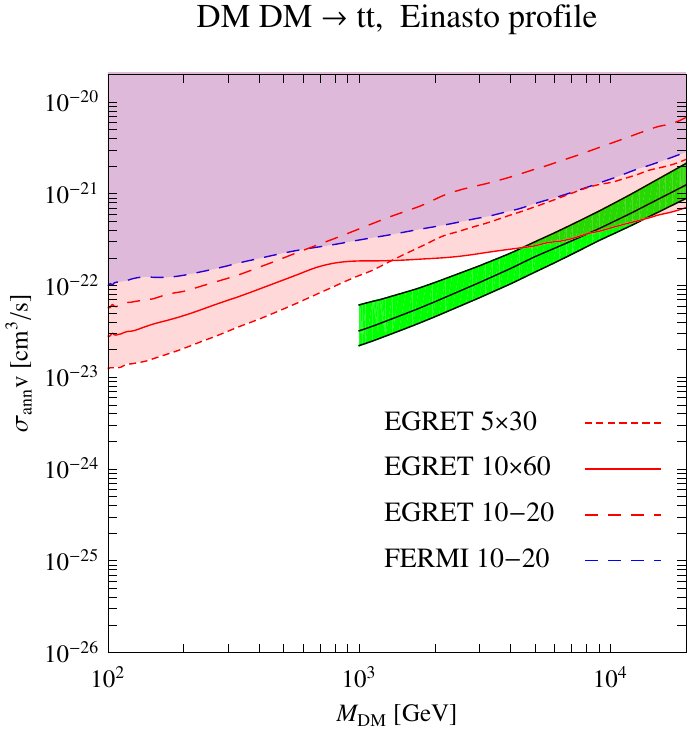}\\[2mm]
\hspace{-8mm}
\includegraphics[width=0.333\textwidth]{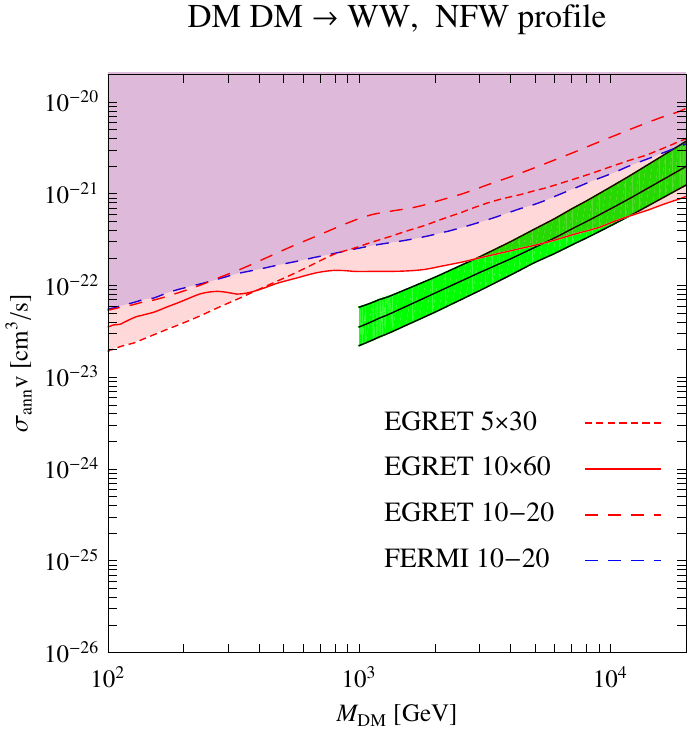}\
\includegraphics[width=0.333\textwidth]{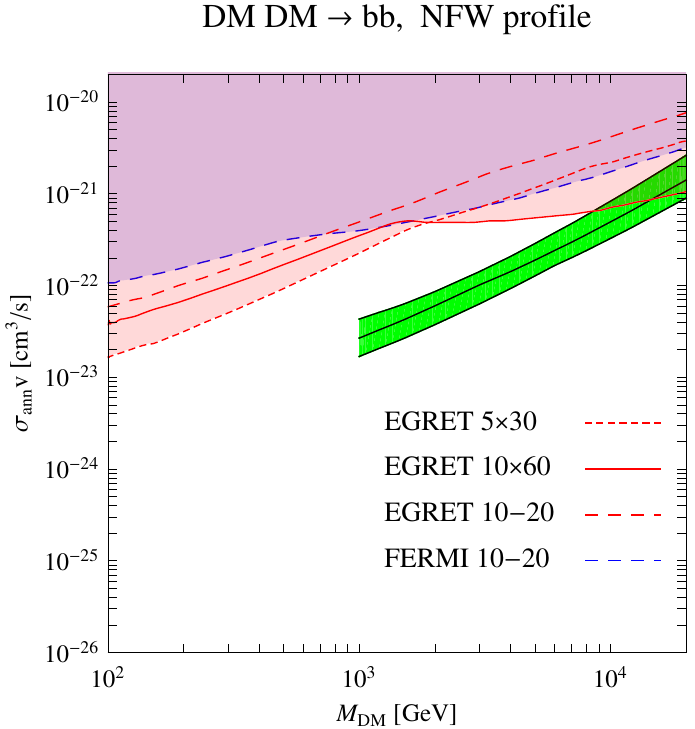}\
\includegraphics[width=0.333\textwidth]{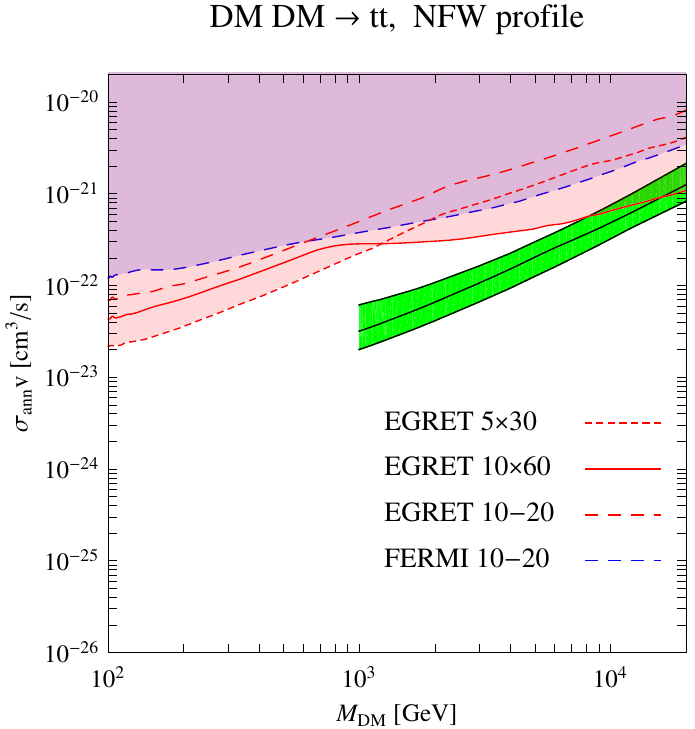}\\[2mm]
\hspace{-8mm}
\includegraphics[width=0.333\textwidth]{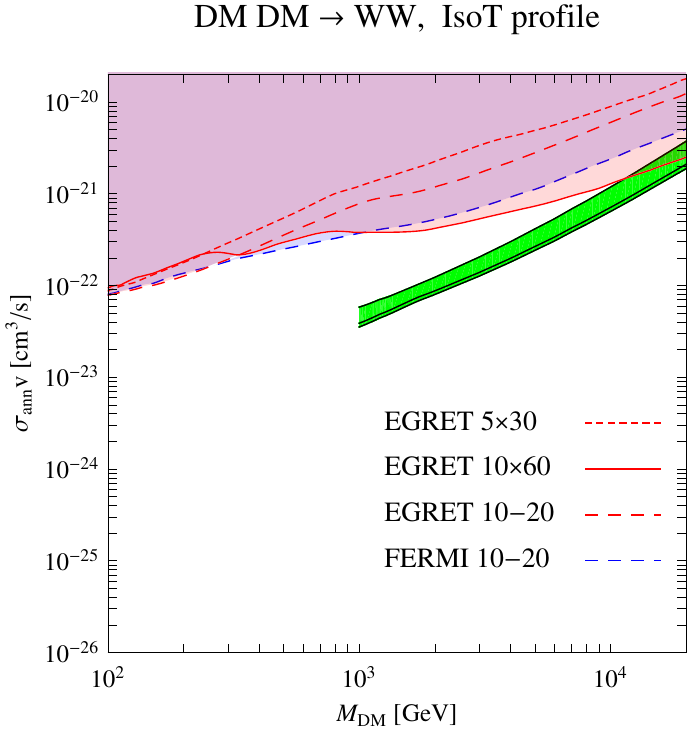}\
\includegraphics[width=0.333\textwidth]{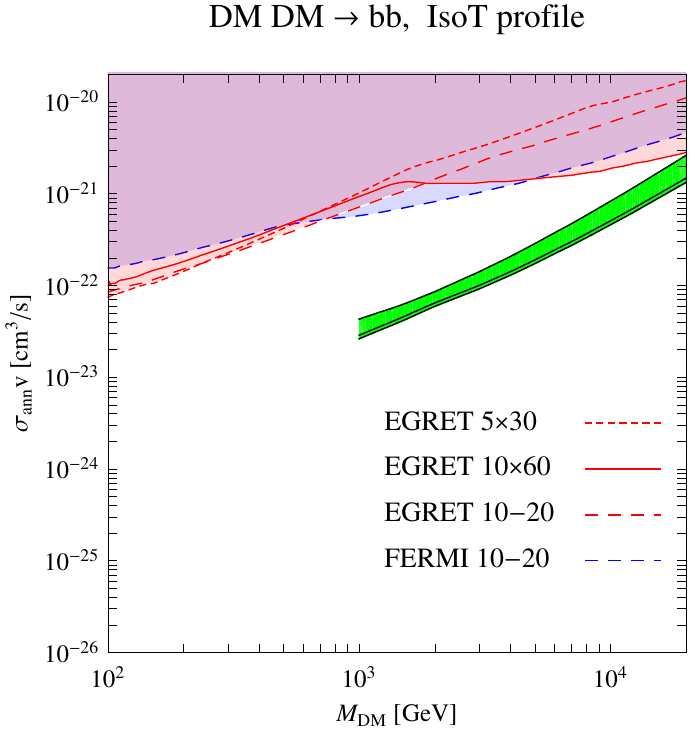}\
\includegraphics[width=0.333\textwidth]{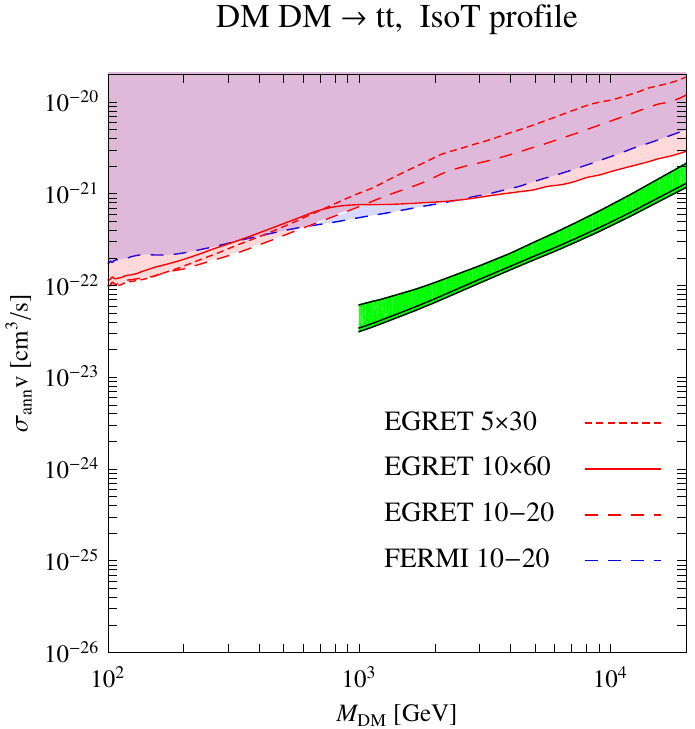}
\caption{\em\label{fig:exclusion2} As in the previous fig.~\ref{fig:exclusion}, but for $W^+W^-$, $b \bar b$ and $t \bar t$ annihilation channels. Since a DM particle fitting the PAMELA data has to be multi-TeV, the green bands start at larger masses. There is no possibility to fit the ATIC data in these channels.}
\end{center}
\end{figure}

In figure\fig{signals} we show some examples of the ICS signals ($\epsilon_1^2\cdot d\Phi/d\epsilon_1\Delta\Omega$) for selected DM models, one for each of the observational regions that we study.\\ 
In fig.\fig{signals}a we plot the ICS gamma ray flux in the `5$\times$30' region, from a DM candidate of $M_{\rm DM}=3$ TeV annihilating into $\tau^+\tau^-$ with $\langle \sigma_{\rm ann}v \rangle = 1.7 \cdot 10^{-22}\, {\rm cm}^3/{\rm sec}$. We have assumed an Einasto DM profile. We plot the individual contributions of ICS signals onto the CMB (green), IR (red) and SL (blue) photon fields, and the resulting total flux (black thicker). As we see, the total flux skims through the EGRET data points at the largest energies. Our result compares with the result in ref.~\cite{700+}, obtained with GALPROP, modulo a slightly different choice for the Einasto profile and possibly the precise definition of the observation region. We find however a good agreement.\\
In fig.\fig{signals}b we plot the ICS gamma ray flux in the `10$\times$60' region, from a DM candidate of $M_{\rm DM}=1.5$ TeV annihilating into $\mu^+\mu^-$ with $\langle \sigma_{\rm ann}v \rangle = 5 \cdot 10^{-23}\, {\rm cm}^3/{\rm sec}$. We have assumed an isothermal DM profile. Here as well the flux lies just below the EGRET data points. Our result here compares with the result in ref.~\cite{Cuoco}, obtained with GALPROP, modulo a slightly different choice for the isothermal profile. We again find however a good agreement between our computation and the fully numerical one.\\
Finally, in fig.\fig{signals}c we plot the ICS gamma ray flux in the `10$-$20' strips, from a DM candidate of $M_{\rm DM}=10$ TeV annihilating into $W^+W^-$ with $\langle \sigma_{\rm ann}v \rangle = 5 \cdot 10^{-22}\, {\rm cm}^3/{\rm sec}$,\footnote{These parameters are characteristic of the Minimal Dark Matter model, see e.g.~\cite{MDM}.} assuming an NFW DM profile. The flux here lies well below the EGRET and FERMI datapoints. 

The comparisons with the full GALPROP results also allows us to validate the approximations of our semi-analytical treatment. In particular, the diffusion processes that we have neglected can redistribute somewhat the population of $e^\pm$ that originate the ICS signal. In the case of the isothermal profile the effect is expected to be minor, as the DM density is essentially uniform over a broad central region of the galaxy. In the case of peaked profiles (Einasto, NFW) the $e^\pm$ can be moved from the denser central regions towards the outside. This would slightly reduce the signal from observational regions centered around the GC (such as the `$5 \times 30$' and `$10 \times 60$' regions) and instead increase it for regions outside the GC (such as the `10$-$20' strip). 
Comparing with the fully numerical results presented in~\cite{700+,Cuoco}, we estimate at about a factor of 2 the systematic error introduced by our approximations \footnote{{\it Note added:} Ref.~\cite{StrumiaPapucci} also in fact finds that the diffusion-less treatment produces fluxes that are good within a factor of two over the whole photon energy range in the `10-20' region, see their fig.4.}.\label{discussiondiffusion}

We are only including the ICS contribution in these plots and in our work, and not the prompt $\gamma$-rays from DM annihilations. The latter ones are model dependent and are expected to be subdominant in these low average DM density regions, especially for large DM masses.

\medskip

Scanning systematically a large range of DM masses $M_{\rm DM}$ (100 GeV - 20 TeV) and considering all the annihilation channels of eq.\eq{channels}, we derive then constraints on the annihilation cross section by the  following conservative prescription. For each observational region with the corresponding data points, fixed an annihilation channel, a DM distribution profile and a DM mass, we impose that the ICS signal must not exceed any of the experimental data points by more than 3$\sigma$. This determines a maximum annihilation cross section. Notice that this is the most conservative possible approach, as we do not assume anything on the background. An alternative procedure would be to take the EGRET and FERMI signals to be of astrophysical origin and then impose the requirement that the signal from Dark Matter, summed to such astrophysical background, does not exceed the data points. This would make our final constraints much stronger. Since we do not want, however, to exclude more portions of the parameter space that it is justified by current data, we stick to the first prescription. 

\medskip

Figure\fig{exclusion} and\fig{exclusion2} shows our results. We superimpose the ICS constraints to the regions identified by the PAMELA data (green bands, the different lines represent different positron and electron propagation models). We also indicate the subregions that allow to fit the ATIC peak as well (in yellow), see~\cite{CKRS,BCST} for the whole analysis.

Focussing first on the purely leptonic annihilation channels (fig.\fig{exclusion}), one sees that, choosing the Einasto profile, the ICS bounds exclude a large portion of the PAMELA regions.  
In general, the most constraining bound is the one from the EGRET data points in the `10$\times$60' region, both because they are characterized by smaller error bars and because their spectrum falls rapidly at large energies (see fig.\fig{signals}b).

It is worth pointing out that the exclusion lines of the $`10-20'$ strips increase more steeply at large masses with respect to the $5\times30$ and $10\times60$ region ones. This is due to the fact that the data in such regions go up to 80 GeV (see figs.\fig{signals}a and\fig{signals}b), whereas in the $`10-20'$ strips they reach only 6 GeV for EGRET and 9 GeV for FERMI (see fig.\fig{signals}c) and thus the fluxes are less constrained. One expects that, when FERMI will provide data in all the regions of the sky and cover a bigger range of energy, it will have the power to probe the parameter space down to much smaller DM annihilation cross sections. If the EGRET `GeV excess' will be ruled out in the other regions as well, this actually means that it can be expected that the exclusion lines from FERMI will become very stringent at large masses.

One also sees that the ${\rm DM}\, {\rm DM} \to e^+e^-$ channel is generically more constrained than $\mu^+\mu^-$ or $\tau^+\tau^-$, something which has to do with the fact that the original $e^+e^-$ population from such channel is concentrated at larger energies, and therefore produces higher energy ICS photons, where the datapoints are more constraining.

\smallskip

Assuming different DM profiles relaxes slightly the bounds. For an isothermal profile, some portion around $M_{\rm DM} = 1\, {\rm TeV}$ are reopened. We notice that the Einasto profile leads to the most stringent bounds: this is just because such a profile predicts a larger DM density at radii around several kpc (see fig.\fig{ISRF}a), which is where the regions that we are considering are located. The fact that the density prescribed by the NFW profile becomes larger  than Einasto very close to the Galactic Center is not relevant for these broad field integrated signals.
We also recall that the bounds relative to the isothermal profile are not expected to be affected by our no-diffusion approximation, so they can be considered as the most robust. 

\medskip

Fig.\fig{exclusion2} shows that the same qualitative features are present in the `hadronic' annihilation (including $W^+W^-$) channels. Notice that the ICS bounds are only a factor of $\mathcal{O}$(few) higher than the corresponding purely leptonic ones, but the PAMELA regions are less constrained, because they are located at smaller cross section values (especially for large masses) for these channels.\footnote{At small masses, the exclusion line from the preliminary FERMI datapoints in the `10$-$20' strip becomes less stringent than the corresponding line of EGRET. This is due to the fact that, for the `hadronic' annihilation channels, the primary electron and positron spectra, and therefore the ICS $\gamma$-ray spectra, are broader, and in particular are sizable at {\em low} energy, where EGRET has currently more points than those available in the preliminary FERMI release that we are using.}

\medskip

We specify that we are always assuming unit boost factor and no other particle physics enhancement (such as the Sommerfeld or resonance enhancement). If such factors are present, then it simply means that the quantity constrained on the vertical axis of our panels in fig.\fig{exclusion} and\fig{exclusion2} is the product $B \times \sigma v$ or $S \times \sigma v$ (where $B$ denotes the astrophysical boost factor and $S$ an enhancement): the PAMELA regions and the ICS contours simply rescale together, as the effect responsible for the increase would be most probably present in both cases (the local galactic halo from which the PAMELA positrons come and the halo regions probed by EGRET and FERMI in gamma rays) in the same way. An exception to this is the possibility that a single local DM clump is responsible for the PAMELA signal; in that case the ICS signals from other regions of the halo would not contain the effect. This possibility looks however rather unlikely~\cite{localsubhalo}. A detailed analysis of the distribution of substructures in the halo, and of the DM velocity dispersion, would be necessary to go beyond these qualitative statements (see e.g.~\cite{bovy}).

\section{Conclusions}
\label{concl}

We have studied the compatibility of the interpretation in terms of DM annihilations of the excesses of electrons and positrons, recently claimed by CR experiments such as PAMELA and ATIC, with the observation of diffuse gamma ray fluxes, as they would be produced by Inverse Compton scatterings on interstellar photons of such large populations of energetic $e^\pm$ in the galactic halo. We first presented a review of the basic formalism needed to compute the ICS signal in the context of DM annihilation, and then discussed its concrete application. We have performed the computation of the expected ICS fluxes semi-analytically (as opposed to running numerical galactic codes).
We have worked under the simplifying assumption of neglecting the diffusion of the electrons away from their production spot in the galactic halo.
Moreover, we have adopted simplifying modelizations of the InterStellar Radiation Field, based however on the precise computations in the literature. We find good agreement between the results of our procedure and the fully numerical computations. We can quantify at about a factor of 2 the systematic error introduced by our approximations.

We have considered three main regions of observation and compared the expected fluxes with the existing EGRET data and with the preliminary FERMI points, deriving constraints on the plane $M_{\rm DM}/\sigma_{\rm ann}v$.
The figures\fig{exclusion} and\fig{exclusion2} summarize our results. They show that, for purely leptonic annihilation channels, large portions of the PAMELA regions at $M_{\rm DM}$ bigger than a few hundred GeV are excluded by a factor ranging from `a few' up to about one order of magnitude, if an Einasto DM profile is assumed (for an NFW profile a very similar situation holds). This includes in particular the $M_{\rm DM} \approx 1$ TeV region. If an isothermal profile (disfavored however by numerical simulations) is chosen, the constraints are somewhat relaxed, but still most of the $M_{\rm DM} >$ few TeV area is excluded. For the hadronic and $W^+W^-$ annihilation channels the qualitative results are similar, but larger portions of the PAMELA regions remain allowed.

\medskip

Complementary constraints on DM annihilation are imposed by high energy gamma rays (generated directly from the DM annihilation process) from the galactic center region and from satellite galaxies and by synchrotron radiation (generated by $e^\pm$ in the galactic center's magnetic field). A model-independent analysis, analogous to the one that we performed here, has been carried out in~\cite{BCST}. Its results show that the regions of the parameter space that allow to fit the PAMELA (and ATIC) data are disfavored by about one order of magnitude if a benchmark Einasto or NFW profile is assumed.
But choosing a smoother profile and/or assuming that a part of the cross section is due to an astrophysical boost factor that would not be present in dwarf galaxies and the Galactic Center due to tidal disruption re-allows part of the space.\\ 
The synchrotron and gamma ray constraints are therefore qualitatively quite similar and quantitatively somewhat stronger than those imposed by ICS, studied in this paper. They are however apparently quite dependent on the choices of the DM distribution, as expected from the fact that they come mainly from signals originating at the Galactic Center. On the contrary, the ICS signals (and therefore the constraints) are more robust with respect to these DM-distribution related issues. 

\bigskip

To conclude, our results show that the ICS gamma ray fluxes have the power of excluding large portions of the parameter space (in DM mass and annihilation cross section) individuated by the PAMELA data, especially for masses larger than about 1 TeV, for purely leptonic annihilation channels and for benchmark Einasto or NFW profiles. This adds tension to the DM interpretation of the PAMELA data, on top of the already existing constraints from prompt high energy gamma rays and synchrotron radiation from the same DM annihilations~\cite{BCST}: gamma rays seem to disfavor the DM interpretation of the PAMELA data in a {\em multiwavelength} range. Future results from the FERMI satellite will have the power to probe larger portions of the parameter space.

\paragraph{Acknowledgements}
We thank Alessandro Cuoco and Michele Papucci for discussions and Nicola Giglietto for useful information. We don't know how to thank Alessandro Strumia for so many useful discussions, comments and confrontations.

The work of P.P. is supported in part by the International Doctorate on AstroParticle Physics (IDAPP) program. 
We thank the EU Marie Curie Research \& Training network ``UniverseNet" (MRTN-CT-2006-035863) for support.

\bigskip
\appendix

\footnotesize
\begin{multicols}{2}
  
\end{multicols}


\begin{thebibliography}{nn}


\bibitem{reviews}
Recent reviews include: 
\art{G. Jungman, M. Kamionkowski, K. Griest}{Phys. Rep.}{267}{195}{1996}.
G.~Bertone, D.~Hooper and J.~Silk,
  Phys.\ Rept.\  405 (2005) 279
  [arXiv:hep-ph/0404175].
J.~Einasto,
  arXiv:0901.0632 [astro-ph.CO].

\bibitem{PAMELA}
 P.~Picozza {\it et al.},
  Astropart.\ Phys.\  {\bf 27} (2007) 296
  [arXiv:astro-ph/0608697].

\bibitem{PAMELApositrons}
O.~Adriani {\it et al.}  [PAMELA Collaboration],
  arXiv:0810.4995.

\bibitem{HEAT}
S.~W.~Barwick {\it et al.}  [HEAT Collaboration],
  Astrophys.\ J.\  482 (1997) L191
  [arXiv:astro-ph/9703192].

\bibitem{AMS01}
AMS-01 Collaboration: \art[astro-ph/0703154]{M. Aguilar et al.}{Phys. Lett.}{B646}{145-154}{2007}.

 \bibitem{PAMELApbar}
 O.~Adriani {\it et al.},
  arXiv:0810.4994.

\bibitem{CKRS}
M.~Cirelli, M.~Kadastik, M.~Raidal and A.~Strumia,
  Nuclear Physics B 813 (2009), pp. 1-21
  [arXiv:0809.2409].

\bibitem{resonance}
D.~Feldman, Z.~Liu and P.~Nath,
  Phys.\ Rev.\  D {\bf 79} (2009) 063509
  [arXiv:0810.5762 [hep-ph]].
  M.~Ibe, H.~Murayama and T.~T.~Yanagida,
  arXiv:0812.0072 [hep-ph].
  W.~L.~Guo and Y.~L.~Wu,
  Phys.\ Rev.\  D {\bf 79} (2009) 055012
  [arXiv:0901.1450 [hep-ph]].

  \bibitem{Sommerfeld}
A. Sommerfeld, ``\"Uber die Beugung und Bremsung der Elektronen'', Ann. Phys. 403, 257 (1931).
J.~Hisano, S.~Matsumoto and M.~M.~Nojiri,
  Phys.\ Rev.\ Lett.\  {92} (2004) 031303
  [arXiv: hep-ph/0307216].
J.~Hisano, S.~Matsumoto, M.~M.~Nojiri and O.~Saito,
  Phys.\ Rev.\  D {71} (2005) 063528
  [arXiv: hep-ph/0412403].
See also previous work in 
  K.~Belotsky, D.~Fargion, M.~Khlopov and R.~V.~Konoplich,
  Phys.\ Atom.\ Nucl.\  {71} (2008) 147
  [arXiv:hep-ph/0411093] and references therein. 

\bibitem{MDMastro}
M.~Cirelli, A.~Strumia, M.~Tamburini,
  Nucl.\ Phys.\  B {\bf 787} (2007) 152
  [arXiv:0706.4071 [hep-ph]].

\bibitem{Arkani}
N.~Arkani-Hamed, D.~P.~Finkbeiner, T.~R.~Slatyer and N.~Weiner,
  Phys.\ Rev.\  D {\bf 79} (2009) 015014
  [arXiv:0810.0713 [hep-ph]].

\bibitem{Sommerfeld2}
 M.~Lattanzi and J.~I.~Silk,
  arXiv:0812.0360 [astro-ph].
  J.~D.~March-Russell and S.~M.~West,
  arXiv:0812.0559 [astro-ph].
L.~Pieri, M.~Lattanzi and J.~Silk,
  arXiv:0902.4330 [astro-ph.HE].
  R.~Iengo,
  arXiv:0902.0688 [hep-ph].

\bibitem{Lavalle}
See, for a recent analysis, \hepart[astro-ph/0603796]{J.~Lavalle, J.~Pochon, P.~Salati and R.~Taillet}
and 
\hepart[0709.3634]{J.~Lavalle, Q.~Yuan, D.~Maurin and X.~J.~Bi}.
See also 
F.~Donato, D.~Maurin, P.~Brun, T.~Delahaye and P.~Salati,
  Phys.\ Rev.\ Lett.\  {\bf 102} (2009) 071301
  [arXiv:0810.5292 [astro-ph]].

\bibitem{bovy}
J.~Bovy,
  arXiv:0903.0413 [astro-ph.HE].

  \bibitem{ATIC-2}
ATIC collaboration, Nature 456 (2008) 362.

\bibitem{PPB-BETS}
\hepart[0809.0760]{PPB-BETS collaboration}. Web page: \url{http://ppb.nipr.ac.jp}.

\bibitem{HESSleptons}
 F.~Aharonian {\it et al.}  [H.E.S.S. Collaboration],
  Phys.\ Rev.\ Lett.\  101 (2008) 261104
  [arXiv:0811.3894].

\bibitem{FERMIleptons}
{\sc FERMI} collaboration, `{\em Measuring 10-1000 GeV Cosmic Ray Electrons with GLAST/LAT}',
talk at the ICRC07 conference.

 \bibitem{pulsars}
A.~M.~Atoian, F.~A.~Aharonian and H.~J.~Volk,
Phys.\ Rev.\  D {52} (1995) 3265.
\hepart[0804.0220]{I. B\"ushing et al.}.
 T.~Kobayashi, Y.~Komori, K.~Yoshida and J.~Nishimura,  
Astrophys.\ J.\  {601} (2004) 340
[astro-ph/0308470]. See also the recent study in:
D.~Hooper, P.~Blasi and P.~D.~Serpico,
arXiv:0810.1527.
H.~Yuksel, M.~D.~Kistler and T.~Stanev,
  arXiv:0810.2784 [astro-ph].
S.~Profumo,
  arXiv:0812.4457 [astro-ph].
N.~Kawanaka, K.~Ioka and M.~M.~Nojiri,
  arXiv:0903.3782 [astro-ph.HE].
See also P.~D.~Serpico,
arXiv:0810.4846  for an agnostic analysis.

\bibitem{Piran}
N.~J.~Shaviv, E.~Nakar and T.~Piran,
  arXiv:0902.0376 [astro-ph.HE].

\bibitem{Blasi}
P.~Blasi,
  arXiv:0903.2794 [astro-ph.HE].
See also 
  P.~Blasi and P.~D.~Serpico,
  arXiv:0904.0871.
and 
S.~Dado and A.~Dar,
  arXiv:0903.0165 [astro-ph.HE].

\bibitem{exploding}
P.~L.~Biermann, J.~K.~Becker, A.~Meli, W.~Rhode, E.~S.~Seo and T.~Stanev,
  arXiv:0903.4048 [astro-ph.HE].

\bibitem{FERMIdiffuse}
See e.g. the talks of Gudlaugur Johannesson at the XLIV$^{\rm th}$ Rencontres de Moriond on Very High Energy Phenomena in the Universe (La Thuile, 1-8 February 2009), Carmelo Sgr\`o at the GGI Dark Matter conference (Florence, 9-11 February 2009) or Nicola Giglietto at the XLIV$^{\rm th}$ Rencontres de Moriond ElectroWeak (La Thuile, 7-14 March 2009). 

\bibitem{700+}
I.~Cholis, G.~Dobler, D.~P.~Finkbeiner, L.~Goodenough and N.~Weiner,
  arXiv:0811.3641 [astro-ph].

\bibitem{Zhang}
J.~Zhang, X.~J.~Bi, J.~Liu, S.~M.~Liu, P.~f.~Yin, Q.~Yuan and S.~H.~Zhu,
  arXiv:0812.0522 [astro-ph].

\bibitem{Cuoco}
E.~Borriello, A.~Cuoco and G.~Miele,
  arXiv:0903.1852 [astro-ph.GA].

\bibitem{Barger}
V.~Barger, Y.~Gao, W.~Y.~Keung, D.~Marfatia and G.~Shaughnessy,
  arXiv:0904.2001 [hep-ph].

\bibitem{Galprop}
A.~W.~Strong and I.~V.~Moskalenko,
  Astrophys.\ J.\  {\bf 509} (1998) 212
  [arXiv:astro-ph/9807150].
  Webpage: \myurl{galprop.stanford.edu}{galprop.stanford.edu}.

\bibitem{StrumiaPapucci}
See P.~Meade, M.~Papucci, A.~Strumia, T.~Volanski work in progress (to appear) for related work.
{\it Note added after publication:} Appeared as arXiv:0905.0480.

\bibitem{Compton}
A.~H.~Compton,
  Phys.\ Rev.\  {\bf 21} (1923) 483.

\bibitem{reviewICS}
G.~R.~Blumenthal and R.~J.~Gould,
  Rev.\ Mod.\ Phys.\  {\bf 42} (1970) 237.

\bibitem{Rybicki}
G.~Rybicki and A.~P.~Lightman, {\it Radiative Processes in Astrophysics}, J.Wiley \& sons, New York, 1979.

\bibitem{reviewCR}
A.~W.~Strong, I.~V.~Moskalenko and V.~S.~Ptuskin,
  Ann.\ Rev.\ Nucl.\ Part.\ Sci.\  {\bf 57} (2007) 285
  [arXiv:astro-ph/0701517].

\bibitem{UllioRegis}
M.~Regis and P.~Ullio,
  Phys.\ Rev.\  D {\bf 78} (2008) 043505
  [arXiv:0802.0234 [hep-ph]].


\bibitem{Buckley}
L.~Bergstrom, P.~Ullio and J.~H.~Buckley,
  Astropart.\ Phys.\  {\bf 9} (1998) 137
  [arXiv:astro-ph/9712318].

\bibitem{Hunter1997}
S.~D.~Hunter {\it et al.}  [EGRET Collaboration],
  Astrophys.\ J.\  {\bf 481} (1997) 205.
  
\bibitem{Sreekumar1997}  
P.~Sreekumar {\it et al.}  [EGRET Collaboration],
  Astrophys.\ J.\  {\bf 494} (1998) 523
  [arXiv:astro-ph/9709257].

\bibitem{FERMIingen}
See for instance 
  W.~B.~Atwood {\it et al.}  [LAT Collaboration],
  arXiv:0902.1089 [astro-ph.IM].

\bibitem{Strong2004}
A.~W.~Strong, I.~V.~Moskalenko and O.~Reimer,
  Astrophys.\ J.\  {\bf 613} (2004) 962
  [arXiv:astro-ph/0406254].

\bibitem{5x30}
A.~W.~Strong {\it et al.},
  Astron.\ Astrophys.\  {\bf 444} (2005) 495
  [arXiv:astro-ph/0509290].
  
 \bibitem{PYTHIA}
 T.~Sjostrand, S.~Mrenna and P.~Skands,
  Comput.\ Phys.\ Commun.\  {\bf 178} (2008) 852
  [arXiv:0710.3820 [hep-ph]].
  Web page: www.thep.lu.se/~torbjorn/Pythia.html

\bibitem{Papucci}
P.~Meade, M.~Papucci and T.~Volansky,
  arXiv:0901.2925 [hep-ph].

\bibitem{Nomura}
 J.~Mardon, Y.~Nomura, D.~Stolarski and J.~Thaler,
  arXiv:0901.2926 [hep-ph].

\bibitem{BertoneTaosoSweden}
  L.~Bergstrom, G.~Bertone, T.~Bringmann, J.~Edsjo and M.~Taoso,
  arXiv:0812.3895 [astro-ph].

\bibitem{Schwetz}
For the long lived case, see: 
I.~Z.~Rothstein, T.~Schwetz and J.~Zupan,
  arXiv:0903.3116 [astro-ph.HE].

\bibitem{ISRF}
T.~A.~Porter, I.~V.~Moskalenko, A.~W.~Strong, E.~Orlando and L.~Bouchet,
  Astrophys.\ J.\  {\bf 682} (2008) 400
  [arXiv:0804.1774 [astro-ph]].

\bibitem{ISRF2}
T.~A.~Porter and A.~W.~Strong,
  arXiv:astro-ph/0507119.

\bibitem{Graham:2005xx}
  A.~W.~Graham, D.~Merritt, B.~Moore, J.~Diemand and B.~Terzic,
  Astron.\ J.\  {\bf 132} (2006) 2685
  [arXiv:astro-ph/0509417].

\bibitem{Navarro:2008kc}
  J.~F.~Navarro {\it et al.},
  arXiv:0810.1522 [astro-ph].

\bibitem{Navarro:1995iw}
  J.~F.~Navarro, C.~S.~Frenk and S.~D.~M.~White,
  Astrophys.\ J.\  {\bf 462} (1996) 563
  [arXiv:astro-ph/9508025].

\bibitem{Bahcall:1980fb}
  J.~N.~Bahcall and R.~M.~Soneira,
  Astrophys.\ J.\ Suppl.\  {\bf 44}, 73 (1980).\\
See (for a recent re-assessment)  
G.~Gentile, P.~Salucci, U.~Klein, D.~Vergani and P.~Kalberla,
  Mon.\ Not.\ Roy.\ Astron.\ Soc.\  {\bf 351} (2004) 903
  [arXiv:astro-ph/0403154] and 
P.~Salucci, A.~Lapi, C.~Tonini, G.~Gentile, I.~Yegorova and U.~Klein,
  Mon.\ Not.\ Roy.\ Astron.\ Soc.\  {\bf 378} (2007) 41
  [arXiv:astro-ph/0703115].

\bibitem{MDM}
M.~Cirelli, N.~Fornengo and A.~Strumia,
  Nucl.\ Phys.\  B {\bf 753} (2006) 178
  [arXiv:hep-ph/0512090].
M.~Cirelli and A.~Strumia,
  arXiv:0903.3381 [hep-ph].

\bibitem{localsubhalo}
D.~Hooper, A.~Stebbins and K.~M.~Zurek,
  arXiv:0812.3202 [hep-ph].
  P.~Brun, T.~Delahaye, J.~Diemand, S.~Profumo and P.~Salati,
  arXiv:0904.0812.

\bibitem{BCST}
   G.~Bertone, M.~Cirelli, A.~Strumia and M.~Taoso, 
  JCAP03 (2009) 009
  [arXiv:0811.3744].

\bibitem{FERMIleptons2}
{\it Added after publication:}
A.~A.~Abdo {\it et al.}  [The Fermi LAT Collaboration],
  arXiv:0905.0025 [astro-ph.HE].

See also: 
D.~Grasso {\it et al.}  [FERMI-LAT Collaboration],
  arXiv:0905.0636 [astro-ph.HE].

\bibitem{HESSleptons2}
{\it Added after publication:}
F.~Aharonian et al. [H.E.S.S. Collaboration],
  arXiv:0905.0105 [astro-ph.HE].


\end{thebibliography}
\end{document}